\documentclass[acmsmall, authorversion]{acmart}
\AtBeginDocument{%
  }

\setcopyright{cc}
\setcctype{by}
\acmJournal{PACMHCI}
\acmYear{2026} 
\acmVolume{10} 
\acmNumber{2} 
\acmArticle{CSCW037} 
\acmMonth{4} 
\acmDOI{10.1145/3788073}

\usepackage{multirow}
\usepackage[dvipsnames, table, xcdraw]{xcolor}
\usepackage{soul}
\usepackage{colortbl}
\usepackage{verbatim}

\usepackage{amssymb}

\begin{document}

\title{Studying Mobile Spatial Collaboration across Video Calls and Augmented Reality}

\author{Rishi Vanukuru}
\affiliation{%
  \institution{ATLAS Institute, University of Colorado Boulder}
  \city{Boulder}
  \state{CO}
  \country{USA}
}
\email{rishi.vanukuru@colorado.edu}

\author{Krithik Ranjan}
\authornote{Both authors contributed equally to this research.}
\affiliation{%
  \institution{ATLAS Institute, University of Colorado Boulder}
  \city{Boulder}
  \state{CO}
  \country{USA}
}
\email{krithik.ranjan@colorado.edu}

\author{Ada Yi Zhao}
\authornotemark[1]
\affiliation{%
  \institution{ATLAS Institute, University of Colorado Boulder}
  \city{Boulder}
  \state{CO}
  \country{USA}
}
\email{ada.zhao@colorado.edu}

\author{David Lindero}
\affiliation{%
  \institution{Ericsson Research}
  \city{Luleå}
  \state{Norrbotten}
  \country{Sweden}
}
\email{david.lindero@ericsson.com}

\author{Gunilla H. Berndtsson}
\affiliation{%
  \institution{Ericsson Research}
  \city{Stockholm}
  \country{Sweden}
}
\email{gunillabern@gmail.com}

\author{Gregoire Phillips}
\affiliation{%
  \institution{Ericsson Inc.}
  \city{Santa Clara}
  \state{CA}
  \country{USA}
}
\email{greg.phillips@ericsson.com}

\author{Amy Banić}
\affiliation{%
  \institution{University of Wyoming}
  \city{Laramie}
  \state{WY}
  \country{USA}
}
\email{abanic@cs.uwyo.edu}

\author{Mark D. Gross}
\affiliation{%
  \institution{ATLAS Institute, University of Colorado Boulder}
  \city{Boulder}
  \state{CO}
  \country{USA}
}
\email{mdgross@colorado.edu}

\author{Ellen Yi-Luen Do}
\affiliation{%
  \institution{ATLAS Institute, University of Colorado Boulder}
  \city{Boulder}
  \state{CO}
  \country{USA}
}
\email{ellen.do@colorado.edu}

\renewcommand{\shortauthors}{Vanukuru et al.}

\begin{abstract}
  Mobile video calls are widely used to share information about real-world objects and environments with remote collaborators. While these calls provide valuable visual context in real time, the experience of interacting with people and moving around a space is significantly reduced when compared to co-located conversations. Recent work has demonstrated the potential of Mobile Augmented Reality applications to enable more spatial forms of collaboration across distance. To better understand the dynamics of mobile AR collaboration and how this medium compares against the status quo, we conducted a comparative structured observation study to analyze people's perception of space and interaction with remote collaborators across mobile video calls and AR-based calls. Fourteen pairs of participants completed a spatial collaboration task using each medium. Through a mixed-methods analysis of session videos, transcripts, motion logs, post-task exercises, and interviews, we highlight how the choice of medium influences the roles and responsibilities that collaborators take on and the construction of a shared language for coordination. We discuss the importance of spatial reasoning with one's body, how video calls help participants “be on the same page” more directly, and how AR calls enable both onsite and remote collaborators to engage with the space and each other in ways that resemble in-person interaction. Our study offers a nuanced view of the benefits and limitations of both mediums, and we conclude with a discussion of design implications for future systems that integrate mobile video and AR to better support spatial collaboration in its many forms.
\end{abstract}



\received{May 2025}
\received[revised]{November 2025}
\received[accepted]{December 2025}

\maketitle

\section{Introduction}

Mobile video communication is at present the most widely available experience of sharing spaces as if co-located. Calling a family member to see them while conveying wishes on their birthday, showing the layout of a new apartment to a friend, or reaching out to a supervisor for real-time advice on search and maintenance tasks in the field—these are just some of the ways in which people share aspects of themselves and the spaces they inhabit with people at a distance.
Mobile video connects families across distances \cite{ohara2006EverydayPracticesMobile, licoppe2009CollaborativeWorkProducing, procyk2014ExploringVideoStreaming}, supports remote assistance in the real world \cite{brubaker2012FocusingSharedExperiences, jones2022RescueCASTRExploringPhotos, kim2018EffectCollaborationStyles}, and brings mobility to the otherwise static experience of video conferencing via computers and screens. However, mobile video is not without constraints. Prior work has studied issues of asymmetric interaction and misaligned reference frames during mobile video calls \cite{jones2015MechanicsCameraWork, licoppe2009CollaborativeWorkProducing}, and how this affects the ways people integrate this technology into their everyday lives \cite{ohara2006EverydayPracticesMobile}. A key missing piece is the spatial perception and interaction that is the cornerstone of meeting in-person.

The last decade has seen a rise in mobile devices’ ability to support more spatial interactions with digital content, anchored to real environments through Augmented Reality (AR). Mobile AR is used by millions of people as part of games, social media apps, e-commerce, and education \cite{chatzopoulos2017MobileAugmentedReality, arth2015HistoryMobileAugmented}. Recent research has explored how mobile AR can be leveraged to support more spatial forms of remote communication. By tracking the movement of a mobile device in its local environment \cite{stachniss2016SimultaneousLocalizationMapping}, and sharing this information over a network, it is possible to create experiences where remote users can see spatial representations of each other moving around in their respective locations \cite{muller2017RemoteCollaborationMixed,li2022ARCritiqueSupportingRemote}. By integrating networked camera feeds and 3D scanning pipelines into these applications, research groups (including our own) have developed mechanisms for spatial remote collaboration situated in real-world locations, with realistic representations of self \cite{vanukuru2023DualStreamSpatiallySharing, young2020MobileportationNomadicTelepresence}.
These projects position mobile AR as occupying the middle ground between mobile video calls that are widely available but not very spatial, and high-fidelity spatial collaboration using head-worn displays and depth cameras \cite{orts-escolano2016HoloportationVirtual3D, thoravikumaravel2019LokiFacilitatingRemote} that is still far from widespread adoption.

So far, the promise of mobile AR remote communication has been demonstrated through pilot studies in the contexts of crime scene investigations \cite{datcu2016HandheldAugmentedRealitya}, remote laboratory training \cite{villanueva2022ColabARToolkitRemotea}, feedback and critique sessions for physical designs \cite{li2022ARCritiqueSupportingRemote}, and spatial tours of remote environments \cite{young2020MobileportationNomadicTelepresence}. A key step towards realizing the promise of this medium is to conduct an in-depth analysis of the dynamics of mobile AR collaboration, and systematically compare the ways people work together when using mobile AR versus other, more established means of communication. The insights from such an analysis would help provide stronger evidence regarding the usefulness of mobile AR, and shape the design of future systems that aim for more widespread adoption. We take this step by conducting a comparative structured observation study \cite{mackay2025ComparativeStructuredObservation} between mobile video and mobile AR communication. 
In doing so, we were guided by the following research questions:

\begin{enumerate}
    \item How do people perceive and experience shared spaces when mediated by video and AR calls? 
    \item How do people perceive and interact with representations of each other in the real and remotely shared space? 
    \item How does a common understanding of the shared space emerge across local and remote collaborators?
\end{enumerate}

\begin{figure}[h!]
  \includegraphics[width=\textwidth]{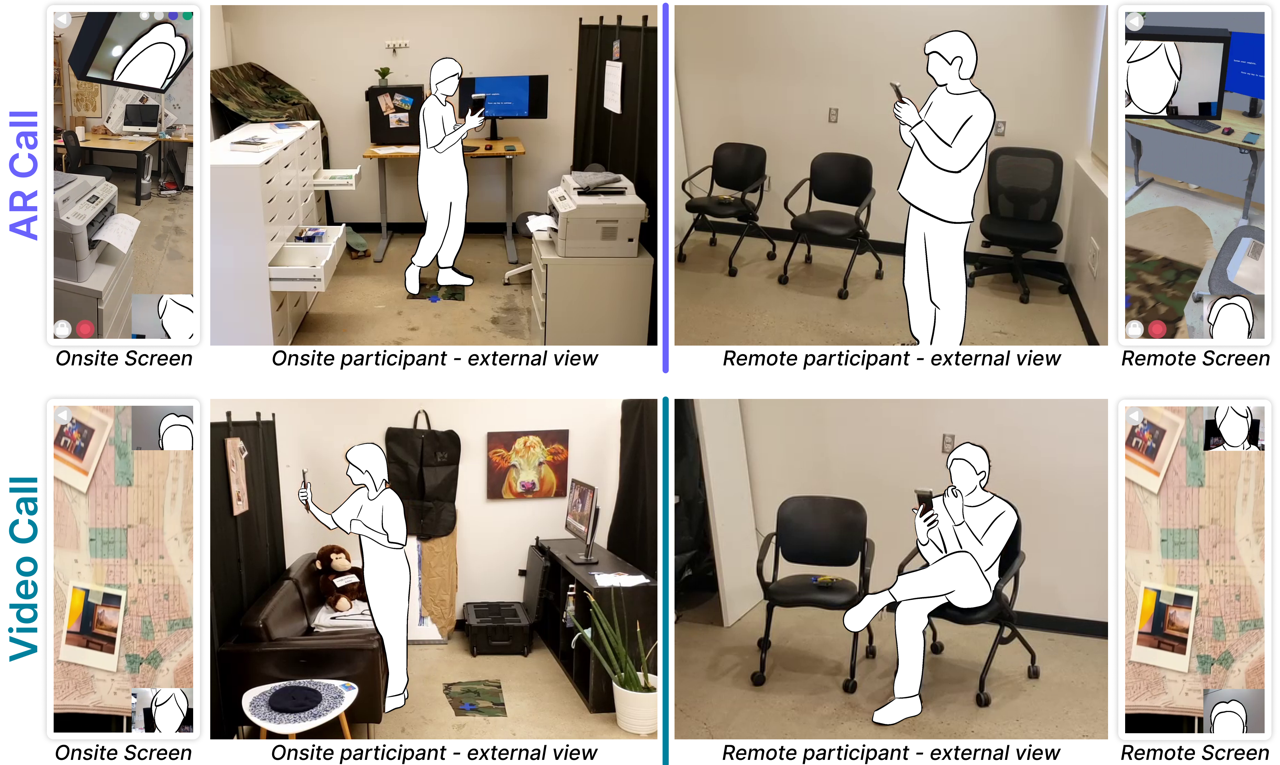}
  \caption{An overview of the spatial collaboration study, using Session 9 as an example. (Top) The AR Call. Onsite Participant 9 (OP9, center-left) is in the real space, and can view Remote Participant 9 (RP9), as a moving video feed in AR (left). RP9 is in a different room (center-right), and can view both the moving video of OP9, and a 3D model of the real space in AR. (Bottom) The video call. Here too, OP9 is in the real space (center-left), and uses their phone to share the video of the room with RP9 who is in a different room (center-right).
  }
  \label{fig:teaser}
  \Description{This is a composite figure depicting images of the experiment setup across video calls and AR calls. While the images are photographs, the participants have been outlined and colored fully in white to preserve anonymity. In the top row, there are four images relating to the AR call: A screenshot from the onsite participant's phone, an external image of the onsite participant in the crime scene, an external image of the remote participant in an empty room, and a screenshot from the remote participant's phone. The crime scene is in the form of a messy office cubicle. We see the onsite and remote participants looking through the phone to see a floating video feed of their counterpart. The remote participant can also see a 3D model of the crime scene which they can walk around and inspect. In the bottom row, there are a similar set of four images (two screens, two external views) of the video call, where we see the onsite participant showing a map on a wall to the remote participant.}
\end{figure}

\noindent
Inspired by prior research that studied the mechanics of camera work in mobile video communication \cite{jones2015MechanicsCameraWork}, we are interested in establishing a similar understanding for mobile AR communication. 
We developed a prototype capable of running both mobile video and AR calls, and had pairs of participants work together on spatial collaborative activities (investigating a crime scene where one participant was physically present at the site, and one participant was remote) through each medium (\autoref{fig:teaser}). The data collected consists of audio and video recordings of the phone screens and external movement of participants, in-app logs of their movement in 3D space, post-task questionnaires and worksheets where participants drew maps of the space from memory, and audio recordings of post-experiment semi-structured interviews.

\noindent
Through our analysis, we discuss how the choice of communication modality affects the way collaborators negotiate their roles and responsibilities, how a shared language for coordination is constructed as participants engage in different forms of communicative work, and how one's body plays an important role in making sense of space and interacting with collaborators in a shared environment. While mobile video calls help collaborators ``be on the same page'' more directly because of a common viewpoint into the shared space, mobile AR communication supports new ways of interacting with people and spaces that come close to replicating the in-person experience. Both video calls and AR calls in their current form have limitations, and our study highlights how a context-based combination of the two might be the best way to design future tools for mobile collaboration. Through this work, we contribute a deeper understanding of remote collaboration via both mobile video and mobile AR calls, informed by a comprehensive user study that analyzes participant experience through their reflections and direct observation, which can be used to design tools that best integrate both modalities to support everyday spatial interactions at a distance.

\section{Background}

\subsection{The growing mobility of video conferencing}

Some of the earliest demonstrations of the promise of video-based remote communication came through the idea of Media Spaces \cite{bly1993MediaSpacesBringing} and their associated implementations in various technology research labs through the 1980's and 1990's. By connecting remote workspaces through a mosaic of cameras, displays, microphones, and speakers, Media Spaces supported both work-oriented and informal forms of communication, along with many unexpected affordances relating to individual and group use \cite{mantei1991ExperiencesUseMedia}. These early prototypes contained many features that are commonplace today---the ability to control the display of one's audio and video, placing greater visual attention on the ``primary speaker'', and manipulating the arrangement of video images to facilitate group discussions. Similarly, studies of their use revealed many issues that still persist in most implementations of video-conferencing, such as the dominance of co-located interactions over remote ones, and issues with coordinating action through gestures and eye contact \cite{sellen1995RemoteConversationsEffects, heath1991DisembodiedConductCommunication}. Parallel efforts to design novel interactions around video-conferencing \cite{ishii1991OpenSharedWorkspace, kuzuoka1992SpatialWorkspaceCollaboration} and develop virtual worlds for more symmetric collaboration \cite{benford1998UnderstandingConstructingShared} resulted in ``fractured ecologies'' of interfaces that separated conduct \textit{``from the environment in which it is produced and from the environment in which is received''} \cite{luff2003FracturedEcologiesCreating}. At the same time, the growing understanding of the importance of mobility in Computer Supported Cooperative Work (CSCW) \cite{luff1998MobilityCollaboration} pointed to the potential of mobile video to address some of these challenges. This potential began to be realized in the early 2000's with the rising use of personal mobile phones, and 3G networks that could support basic video telephony.

\citet{ohara2006EverydayPracticesMobile} conducted one of the earliest studies of mobile video use, discussing the reasons for video calling (maintaining social and emotional connections, showing things to talk about, functional talk), and the social and technological barriers to wider use at the time. They noted how mobile video calls \textit{``exist within a complimentary [sic] suite of mobile communication possibilities''}, and called for further design that connected interactions across these possibilities. A key aspect of mobile video calls is the control and responsibility that one has when sharing and framing views for remote inspection. By highlighting the collaborative work that goes into producing ``meaningful'' views, \citet{licoppe2009CollaborativeWorkProducing} present a rich account of everyday mobile video use, at a time when its uptake was still very limited. In the years since, developments in technology and social acceptance have resulted in mobile video finding its way into numerous contexts \cite{brubaker2012FocusingSharedExperiences}, including connecting families across distance \cite{procyk2014ExploringVideoStreaming, inkpen2013Experiences2GoSharingKids} and remote assistance \cite{jones2022RescueCASTRExploringPhotos, unver2018HandsFreeRemoteCollaboration}. While most of these implementations discuss the many benefits of mobile video alongside the functional challenges that emerge, \citet{jones2015MechanicsCameraWork} conducted the first study of the ``mechanics'' of camera work, and discussed how the asymmetries of control, participation, and awareness led to many barriers in communication between people interacting across distance.
Ultimately, the lack of a shared task and reference space \cite{buxton2009MediaspaceMeaningspaceMeetingspace} for sense-making and deictic gestures means that mobile video is constantly at a disadvantage when compared to in-person interaction.

\subsection{Spatial mobile experiences with Augmented Reality}

Augmented and Virtual Reality approaches to remote collaboration have many of the same goals as video-conferencing, but move the domain of interaction away from screens and into space. Early research projects explored the use of AR overlays of people and content in real environments \cite{billinghurst2002CollaborativeAugmentedReality}, and the creation of mixed-reality shared spaces that bridged the real and the virtual \cite{benford1998UnderstandingConstructingShared}. In the years since, these attempts to replicate in-person interactions have led to approaches that use volumetric capture and high-fidelity head-mounted displays to stream real-time representations of people and spaces for remote collaboration \cite{thoravikumaravel2019LokiFacilitatingRemote, orts-escolano2016HoloportationVirtual3D, irlitti2023VolumetricMixedReality, gunkel2021VRCommEndtoendWeb, piumsomboon2019ShoulderGiantMultiScale}. Many of these prototypes require complex setups of hardware that cannot be used outside of laboratory environments. While the last decade has seen a significant rise in consumer AR/VR headsets, these are still far from widespread use, and do not yet provide mechanisms for remote communication based on representations of one's real self and environment, usually employing virtual spaces and avatars instead.

As mobile devices have progressed to include on-board motion sensors and high-resolution cameras, it has become possible to create a wide range of collaborative AR experiences. Some of the first explorations of this capability were in the context of distributed collaboration for crime scene investigations \cite{datcu2016HandheldAugmentedRealitya} and remote assistance \cite{gauglitz2012IntegratingPhysicalEnvironment}. Studies have also investigated how people make sense of a shared virtual space via mobile AR and positional landmarks \cite{muller2017RemoteCollaborationMixed, muller2019QualitativeComparisonAugmenteda}. More recently, mobile AR has been used to have discussions around 3D representations of objects and spaces in the context of product design \cite{li2022ARCritiqueSupportingRemote} and remote lab operation \cite{villanueva2022ColabARToolkitRemotea}. With a focus on the real-time sharing of both one's environment and representations of self, projects like MobilePortation \cite{young2020MobileportationNomadicTelepresence} and our work with DualStream \cite{vanukuru2023DualStreamSpatiallySharing} provide immersive, spatial forms of remote communication by establishing a shared reference space consisting of media from both local and remote sites. While all of these projects demonstrate the promise of such forms of interaction, further studies are needed to build a deeper understanding of how people experience mobile AR collaboration, and in particular, how such forms of interaction compare to the status quo of mobile video conferencing.

\section{Studying Mobile Spatial Collaboration}

To help bridge the gap between our understanding of spatial collaboration using mobile video calls and AR communication tools, we conducted a systematic study of these two modalities using the method of Comparative Structured Observation \cite{mackay2025ComparativeStructuredObservation}. 
This is an interventionist, qualitative-first method where users are provided with \textit{``comparable experiences, either with multiple new design variants or with the status quo, which lets them reflect deeply on the advantages and disadvantages of each''} to derive design implications. The method particularly focuses on both the researchers' qualitative observation and the participants' deep reflection on interacting with two or more design variants. 
In defining the method, Mackay and McGrenere provide a checklist of what it takes to be classified as a good Comparative Structured Observation study \cite{mackay2025ComparativeStructuredObservation}, and we structure the description of our methodology using the dimensions of this checklist.

\subsection{Role of comparison}
\subsubsection{Prototype}
In this study, we seek to compare mobile AR calls with mobile video calls in their ability to support spatial collaboration at a distance.
To ensure that this comparison took place on a level playing field, we developed a prototype application capable of supporting both mobile video and mobile AR communication.

\begin{itemize}
    \item \textit{Video call.} Our application's mobile video call mode resembles commercial applications where on-site and remote participants can share views of themselves using the device's front camera and a view of the space in front of them using the rear camera. We enabled the sharing of both self-view and space-view simultaneously, unlike some present-day video call applications that can only do one or the other.
    \item \textit{AR call.} The AR call mode uses the front-facing camera feed and the phone's movement in its local space to create a real-time representation of the local user that appears as a moving video feed in the remote participant's view. The remote participant can also see a pre-scanned 3D model of the on-site environment through AR, which they can walk around and explore, while viewing their collaborator move in space.
\end{itemize}

\begin{figure}[h!]
  \includegraphics[width=\textwidth]{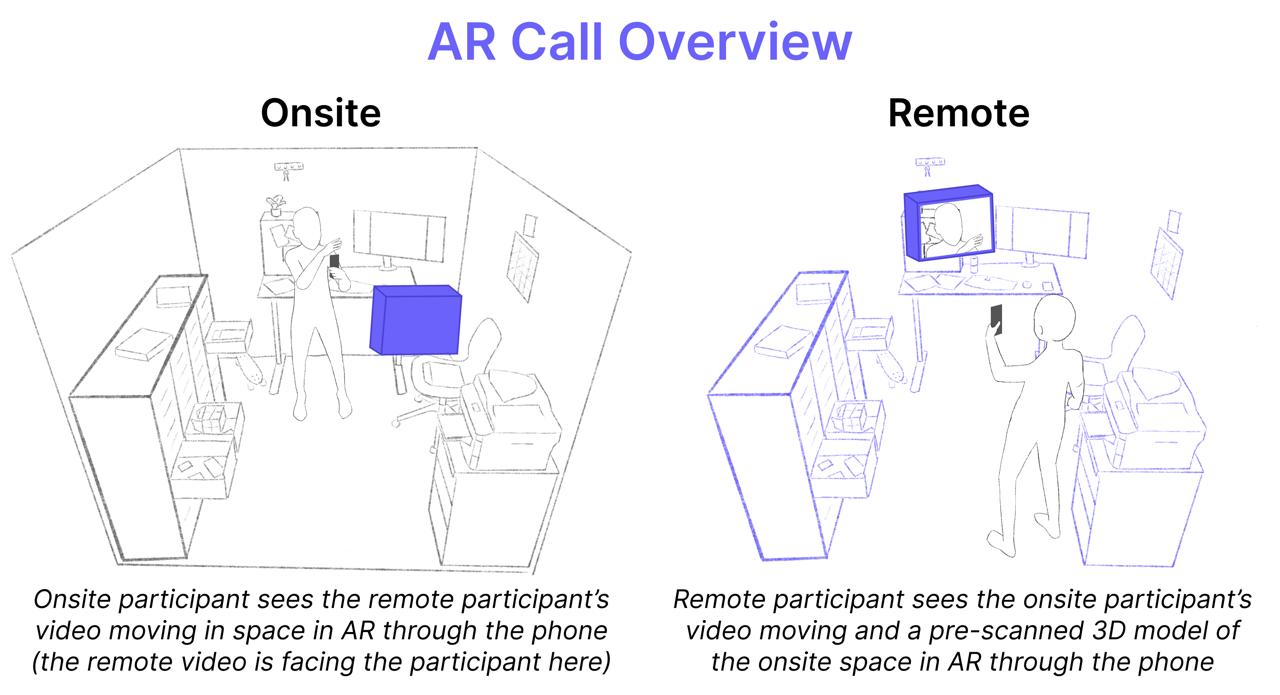}
  \caption{An overview of the AR Call mode. Elements viewed in AR through the phone are colored blue. (Left) The onsite participant is in the real space, and can view the moving video representation of the remote participant in AR through their phone. (Right) The remote participant uses their phone to view a pre-scanned 3D model of the real space, as well as the moving video representation of the onsite participant.}
  \label{fig:ar-call-overview}
  \Description{This figure is composed of two sub-figures arranged left to right. On the left, we see a perspective view of the sketch of one of the crime scenes used in the study, with the onsite participant facing towards us and looking through the phone at a blue rectangle. The caption underneath reads - onsite participant sees the remote participant's video moving in space in AR through the phone (the remote video is facing the participant here). On the right, we see a similar sketch of the room from the same perspective with the remote participant looking away from us at the onsite participant's video through the phone. The caption underneath reads - Remote participant sees the onsite participant's video moving and a pre-scanned 3D model of the onsite space in AR through the phone.}
\end{figure}

\noindent
While prior work in Mobileportation \cite{young2020MobileportationNomadicTelepresence} and DualStream \cite{vanukuru2023DualStreamSpatiallySharing} included features to create and share a 3D scan of spaces in real-time, we instead use a pre-scanned 3D model of the experiment setups to control for visual quality. \autoref{fig:ar-call-overview} illustrates the interaction enabled by the AR call application, and the supplementary video figure\footnote{Supplementary materials available at: \href{https://rishivanukuru.com/studying-mobile-spatial-collaboration}{rishivanukuru.com/studying-mobile-spatial-collaboration}} showcases excerpts from real study sessions, demonstrating the use of video calls and AR calls.

Building a custom prototype enabled us to embed spatial logging mechanisms in both video and AR mode. Further, since most mobile devices do not allow applications to simultaneously access the rear-facing and front-facing camera feeds while also performing AR tracking, our application uses an externally-mounted front-facing camera connected to the phone via USB (similar to prior work \cite{young2020MobileportationNomadicTelepresence, vanukuru2023DualStreamSpatiallySharing}). We built the prototype application for Android devices using Unity\footnote{Unity, from \href{https://unity.com/}{unity.com}}. For the experiment, we used two Samsung Galaxy S20+ phones, each with a RealSense D435\footnote{RealSense D435, from \href{https://www.realsenseai.com/products/stereo-depth-camera-d435/}{realsenseai.com} (originally Intel)} camera to capture the user's self view. Audio and video streaming took place over Agora cloud servers\footnote{Agora, from \href{https://www.agora.io/en/}{agora.io}}, and positional networking was achieved using Photon Fusion\footnote{Photon Fusion, from \href{https://www.photonengine.com/fusion}{photonengine.com}}.

\subsubsection{Participant activities} 
We aimed to study joint spatial exploration behaviors where a local user shares their environment with a remote collaborator, and where both users need to make sense of the shared environment and communicate with each other to complete a task. While prior studies have used ``standardized'' collaborative tasks such as verbal discussions, physical puzzle-solving, and staged task assistance \cite{piumsomboon2019ShoulderGiantMultiScale,kim2018EffectCollaborationStyles}, we opt for the more open-ended approach of a spatial investigation. Such forms of interaction are commonly seen in real-life, for example, when people share the view of a newly arranged workplace, seek assistance to find an object in a new environment, or wish to include remote participants in spatial experiences that would otherwise be inaccessible to them. 

\begin{figure}[h!]
  \includegraphics[width=\textwidth]{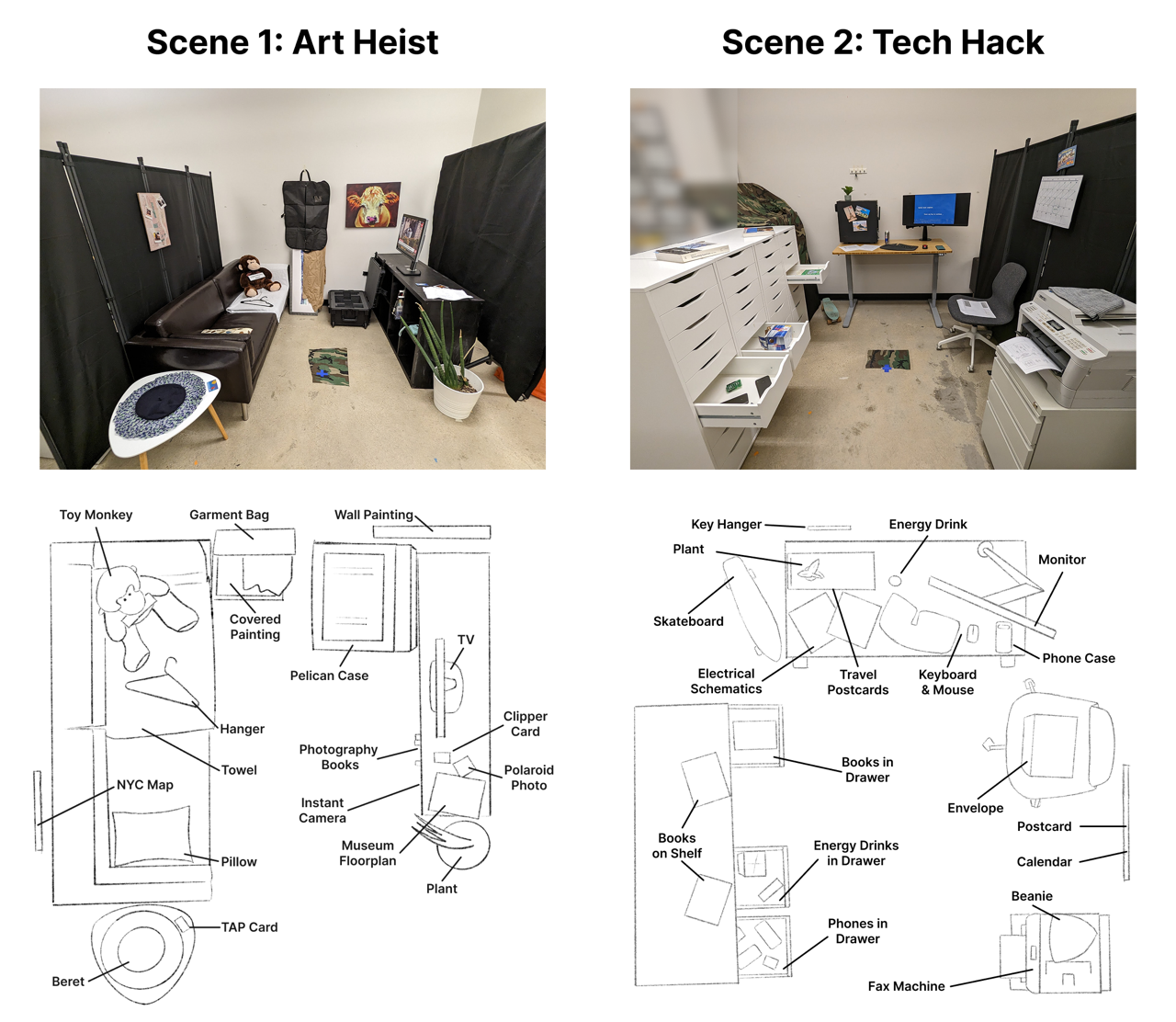}
  \caption{Photographs of the two spatial investigation scenes used in the study, with a corresponding top-down sketch of the space and the various clues spread across the rooms.}
  \label{fig:clues}
  \Description{This figure consists of a 2x2 grid of images. On the top-left, we see a photograph of the Art Heist crime scene, and on the bottom-left, we see a sketch of the top-down view of the Art Heist room with various objects labeled. Going clockwise, these are: (Left) A beret and TAP card on a small side table. A pillow, hanger, and stuffed toy of a monkey all on a couch with a map of New York City on the wall behind it. (Front) A garment bag hanging from a wall above a framed painting leant against the wall and partially covered with paper. A pelican case on the floor nearby, with a hanging canvas painting of a cow above it. (Right) A table with a TV, a Clipper card, polaroid photographs, and a map of a museum on the top. In the shelves under the table, there are books on photography and an instant camera. Next to the table is a medium-sized potted plant. Moving to the next scene, on the top right we see a photograph of the Tech hack room, below which is a similar top-down labeled sketch. The objects, going clockwise, are: (Left) A shelf with multiple drawers. On top of the shelf are two books. A few drawers of the shelf are open and pulled out, containing multiple phones and chips, energy drinks, and another book. (Front) There is a standing desk with a CPU, keyboard, mouse, monitor, and phone case. There are electrical schematics on the desk. There is a small potted plant on top of the CPU, and there are postcards stuck to the front of the CPU. There is a small set of hangers on the wall above the CPU with a set of keys hanging. To the left of the table is a small skateboard on the floor. (Right) A chair with an opened envelope on it, next to which there is a calendar on the wall, with some postcards stuck to it. Next to that is a fax machine on a cabinet with another sheet of schematics coming out, and a beanie on top of it.}
\end{figure}

Building upon prior work which evaluated collaboration via mobile AR using spatial investigations \cite{shakeri2017EscapingTogetherDesign, datcu2016HandheldAugmentedRealitya, poelman2012IfBeingThere}, we constructed two spaces with a series of spatial clues, each relating to an investigation surrounding a different ``crime'' or ``mystery''---an art heist set in a space like an apartment living room, and a technology hack set in a space resembling an office cubicle (\autoref{fig:clues}). These two scenes were created so that we could present different environments for tasks taking place via video call and AR call. We controlled the number, nature, and relative positioning of the clues in both spaces to provide scenes of comparable complexity. In the two task setups, the local participant functions as an ``on-site investigator'' who collaborates with a colleague who is remotely assisting in the case. The on-site participant and the remote participant need to collaborate to make sense of the space and investigate four key questions about the mystery---(1) when the suspect left the site, (2) where they went, (3) what they did at the site, and (4) who they were.
Across both scenes, we designed and arranged the clues to support multiple possible interpretations. As a result, our focus was not on evaluating whether participants found the ``correct'' solution, but rather on whether they were able to discuss the various possibilities and come to a consensus together.

\subsubsection{Experiment Design}
We conducted a within-subjects study, where pairs of participants collaborated remotely using both mobile video and mobile AR to come up with hypotheses regarding the spatial investigation. Experiment sessions lasted between 45 minutes and 1 hour, and began with a researcher briefing the pair of participants about the study, and collecting informed consent. Participants were asked to complete a questionnaire about demographic information and their prior experience with remote collaboration tools. The participants were given a choice between being the onsite or remote investigator, and they remained in their chosen role for both tasks and environments. The participants were then taken to two separate rooms. In the case of the onsite investigator, they were present at the physical crime scene, while the remote investigator was present in an empty area in a different room (\autoref{fig:teaser}). One member of the research team was present at each location. Participants were given a brief description of the crime scene to provide context for the collaborative discussion, and the researchers set up the prototype in either video mode or AR mode. Once the participants began their investigation, they were given around six minutes to explore and discuss, with the researchers providing a time warning when a minute was left. We determined six minutes to be an appropriate time to investigate the scene through pilot tests within the research team. The task was not abruptly stopped at six minutes, and we asked participants to end their discussion at the nearest natural pause in conversation after the allotted time. After this, participants completed a post-task questionnaire and worksheet. This sequence of events was then repeated for the other mode and environment. The order of presentation of modalities and crime scenes was counterbalanced across experiment sessions. At the end of the study session, both participants were brought to the same space, where the researchers conducted a semi-structured interview to understand their subjective experiences across the modalities.

\subsubsection{Recruitment}
Recruitment for the study took place via posters and messages in our university and local community channels and mailing lists. We recruited pairs of participants together to ensure that a lack of familiarity with remote collaborators was not a confounding factor. All recruitment and experimental procedures were approved by the university institutional review board (CU IRB \#24-0086). Each participant received a \$20 gift card as compensation for their time.

\subsubsection{Participant backgrounds}
We recruited 14 pairs of participants (N=28, Mean Age = 25.93, SD = 11.28, Min Age = 18, Max Age = 67) for this study. A majority of these participants were students affiliated with our university, with some participants signing up from the broader community of the area.  
\autoref{tab:descriptive_participants} contains an overview of their responses to the initial questionnaire. We asked questions about their prior experience with video conferencing and mobile video in particular, as well as with head-worn and mobile AR/VR experiences. We see that most participants were quite familiar with video conferencing, and use mobile video calls regularly. Fewer participants use AR/VR headsets in their everyday lives, but most have tried them out to varying extents.

\begin{table}[ht]
    \centering
    \small
    \caption{Participants -- Descriptive statistics from the initial questionnaire}
    \label{tab:descriptive_participants}
    \begin{tabular}{m{3cm} r m{2.1cm} m{2.1cm} m{2.1cm} m{2.1cm}}
        \toprule
        \textbf{Demographics} & \textbf{N} & \textbf{Female} & \textbf{Male} & \textbf{Non-binary} & \textbf{No response} \\
        Gender & 28 & 13 (46.4\%) & 12 (42.9\%) & 2 (7.1\%) & 1 (3.6\%) \\
        \midrule
        \textbf{AR/VR Experience} & \textbf{N} & \textbf{No Experience} & \textbf{Used <5 times} & \textbf{Used between 5 and 10 times} & \textbf{Use frequently (>10 times)} \\
        Head-worn AR & 28 & 15 (53.6\%) & 12 (42.9\%) & 0 (0\%) & 1 (3.6\%) \\
        Head-worn VR & 28 & 2 (7.1\%) & 14 (50\%) & 8 (28.6\%) & 4 (14.3\%) \\ 
        Mobile AR & 28 & 4 (14.3\%) & 11 (39.3\%) & 9 (32.1\%) & 4 (14.3\%) \\ 
        \midrule
        \textbf{Mobile Video Call} & \textbf{N} & \textbf{Rarely} & \textbf{Few times a month} & \textbf{Few times a week} & \textbf{Regularly} \\
        Frequency & 28 & 6 (21.4\%) & 6 (21.4\%) & 6 (21.4\%) & 10 (35.7\%) \\
        \bottomrule
    \end{tabular}
\end{table}

\subsection{Type of Data Collected}
Data about participant experience and reflection was collected through a semi-structured interview conducted after the completion of both tasks. We asked participants to discuss and compare their experience of collaboration across both modalities, and share their hypotheses about each of the crime scenes. During the study tasks, researchers made observations of specific behaviors exhibited by participants and used these as starting points for more detailed discussions about how participants made sense of collaboration across the two modalities. We recorded the audio for each of these interviews. During each task, we recorded the screen view from both phones, the external view of participant movement in both rooms, and audio of their conversation. The prototype also logged the motion of the phone relative to the local space. 
After the completion of each task, participants completed a questionnaire measuring spatial and social presence (questions included in the appendix). They also completed a physical worksheet where they wrote their hypotheses about the mystery, and drew a map of the space from memory (\autoref{fig:drawn-maps} in the appendix).

\subsection{Type of Data Analysis}

\subsubsection{Qualitative analysis of participant reflection and researcher observations}
We used reflexive thematic analysis \cite{braun2006UsingThematicAnalysis, braunvirginia2025ThematicAnalysisPractical} to analyze both the reflections of participants on the design variants, as well as our observations of their interactions during the tasks. Guided by the six key phases of reflexive thematic analysis, we began with an extensive data familiarization exercise, where we manually transcribed each of the task recordings and post-session interviews. We also watched the recordings of the task from the phone screen and external space points of view to create composite and synchronized videos of all four available videos (two screens, two external space recordings) for each task. Three members of the research team then performed a coding exercise on equal subsets of the session and interview data, with a focus on both semantic and latent codes. Once complete, the researchers discussed their codes and underlying analytical process, after which one researcher performed another coding pass over the whole dataset. We used MaxQDA \cite{kuckartz2019AnalyzingQualitativeData} to code both the transcripts and the video recordings. The researchers met multiple times to generate, discuss, and review themes across the two pools of data, the results of which are presented in \autoref{sec:findings}.

\subsubsection{Using quantitative data for context}
The data from the post-task questionnaires about spatial presence were analyzed to provide high-level context for some of the patterns observed in the qualitative discussion. To evaluate the hand-drawn maps as a measure of participant recall, three members of the research team scored each map based on the presence or absence of key objects, and their relative spatial location. The researchers then discussed these ratings to generate a final list of ``recall'' scores for each map.

\subsubsection{Constructing session maps from quantitative data}
To complement the thematic analysis of in-session observations, we sought to summarize some of the spatio-temporal information into visualizations that provided more immediate overviews of key interactions. With the motion logs from the prototype, we were able to create heatmaps and trajectory visualizations of the onsite participant in both tasks, and of the remote participant when they were in the AR call. By operating on both the onsite and remote logs for the AR calls, we created visualizations plotting the distance between participants over time, which allowed us to see moments of intersection with objects in the space, or stable phases of spatial movement. 

\noindent
Using the phone view, overall video recording, and verbal context of communication, we also manually coded for object focus across remote and onsite participants at two levels---object and region (dividing each of the three sides of left, front, and right into two sub-regions). We used the BORIS (Behavioral Observation Research Interactive Software) application to perform this coding \cite{friard2016BORISFreeVersatile}. From here, we were able to generate plots of how participants' individual focus moved across the environment over time, and we overlaid these plots to interpret when participants were thinking and talking about the same part of space.

We collated all these plots to create ``Session Maps'' that helped provide us with visualizations to grasp key spatial and temporal interactions that would otherwise have been challenging to glean through repeated viewings of the recorded videos. 
These session maps were valuable research tools to gain a sense of the nature of collaboration at a glance and assess participant recall, and served as a visual companion during the thematic analysis of in-session interactions. These maps also offer a means of sharing in-session interactions in an anonymized format. We have included sample maps from one session in Figures  \ref{fig:session-spatial} and \ref{fig:session-video} in the Appendix, and further visualizations are included in our supplementary materials page\footnote{Supplementary materials available at: \href{https://rishivanukuru.com/studying-mobile-spatial-collaboration}{rishivanukuru.com/studying-mobile-spatial-collaboration}}.

\section{Vignettes of Collaboration}

To set the stage for our discussion of findings across the various sessions, we begin by highlighting four key examples of the breadth of possible interactions over video and AR. We annotate the onsite participant in session \textit{N} as \textit{OPN}, and the remote participant as \textit{RPN}.

\begin{figure}[ht]
  \centering
  \includegraphics[width=\linewidth]{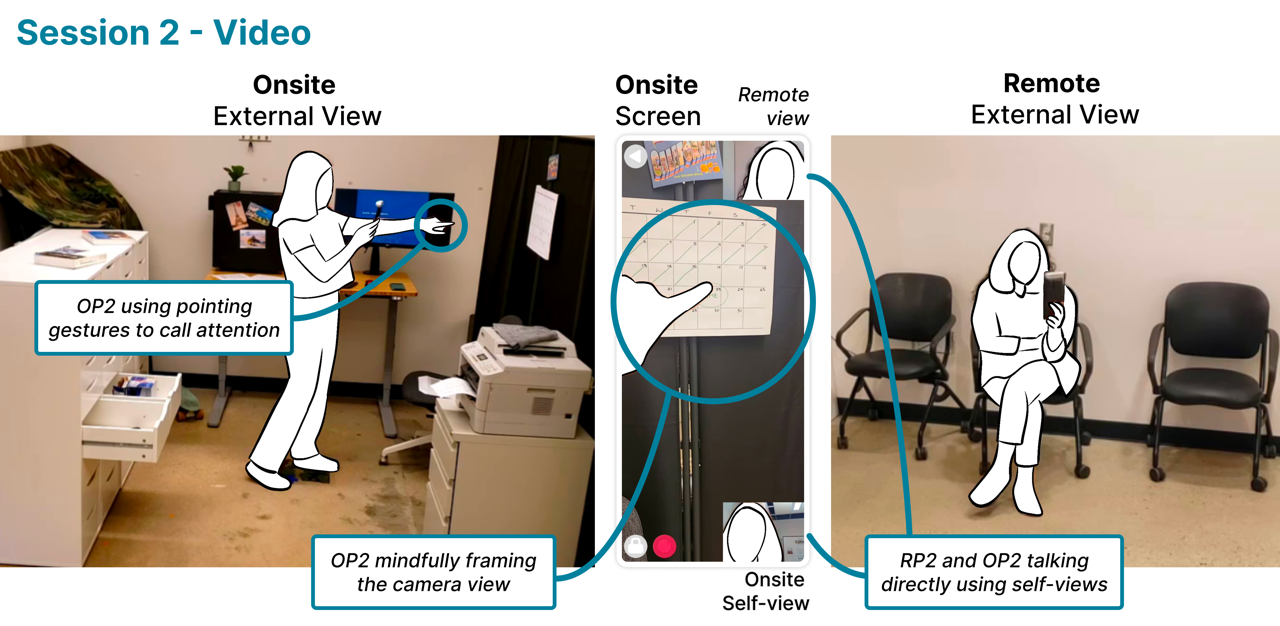}
  \caption{Video call interaction from Session 2. (Left) OP2 using gestures to call attention to objects in the space, while consciously framing the camera view. (Right) RP2 collaborating from a different room.}
    \Description{This figure consists of three images (with people similarly sketched over and anonymized) - an external view of OP2, a screenshot from OP2's screen during the video call, and an external view of RP2. OP2 is using their hand to point at the calendar in the Tech hack scene. On the phone screen, we see the hand pointing to a specific date on the calendar, with both of OP2 and RP2's faces clearly visible in their self video feeds. RP2 is sitting on a chair and holding the phone in one hand.}
  \label{fig:video-vignette-1}
\end{figure}

\paragraph{Session 2 - Video}
In Session 2 (\autoref{fig:video-vignette-1}), OP2 was particularly mindful of the framing of the camera view, ensuring that their own physical view largely corresponded to what was being displayed on the screen. Both OP2 and RP2 made use of their self-video feeds to talk directly to each other, especially when confirming details or discussing hypotheses. While showing the space, OP2 moved between providing overviews and more close-up shots of the clues, all while narrating their movement in the space to provide more context. They also made effective use of pointing gestures through the screen to call RP2's attention to specific objects in their shared visual field. 

\begin{figure}[ht]
  \centering
  \includegraphics[width=\linewidth]{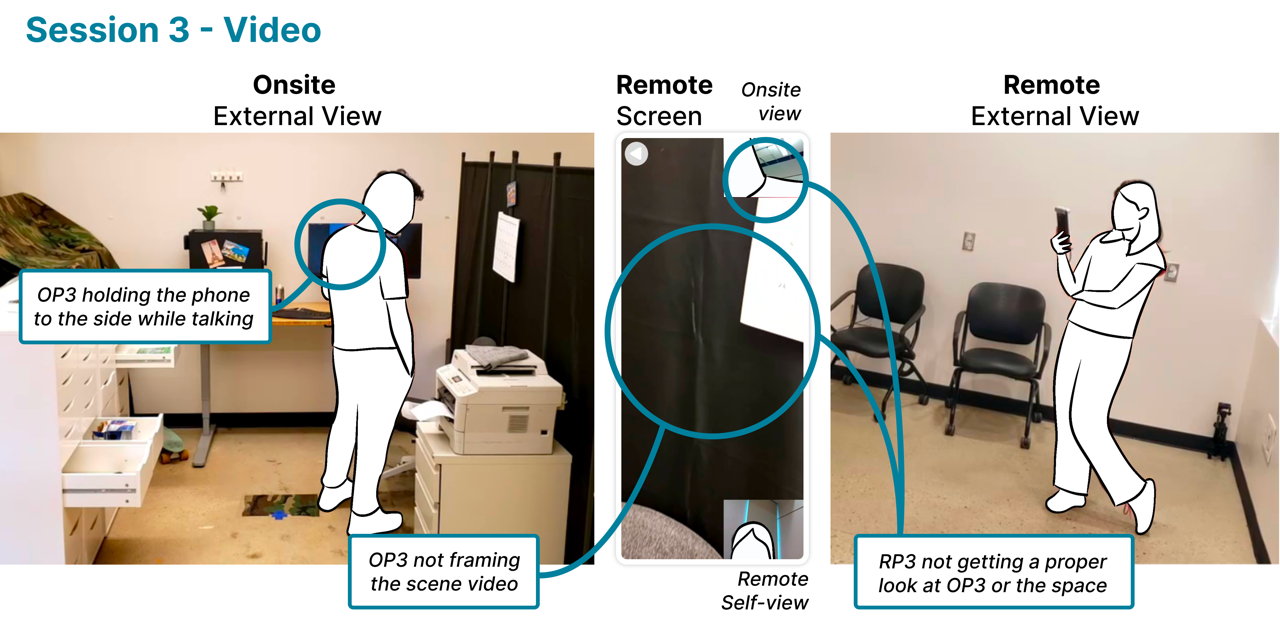}
  \caption{Video call interaction from Session 3. (Left) OP3 is exploring the space and narrating their thoughts while holding the phone to the side. (Center) As a result, both their self-view video and the environment video are not framed properly in the shared feed. (Right) RP3 does not have a proper view into the space, and cannot engage in face-to-face communication.}
  \label{fig:video-vignette-2}
  \Description{This figure consists of three images (with people similarly sketched over and anonymized) - On the left, we see an external view of OP3 in the tech hack scene looking at the calendar while holding the phone away from their face and to the left. In the center, we see a screenshot from RP2's phone, showing that the video is just looking at the wall, and at OP2's neck. On the right, RP2 is standing in the remote room while looking at the phone.}
\end{figure}

\paragraph{Session 3 - Video}
In Session 3 (\autoref{fig:video-vignette-2}), OP3 saw the task of solving the mystery to be more important than working to provide RP3 a full view of the scene. Instead of looking through the phone screen to be mindful of the framing of the shared video, OP3 held the phone to the side, assuming that the point of view would roughly correspond to what they were seeing in the space. As a result, there were many times when RP3 was looking at unframed videos of both the scene and OP3, reducing their collaboration to more audio-based communication. While RP3 would ask questions about objects as they came up on the video, OP3 would not always track back to discuss previously shown objects. 

\begin{figure}[ht]
  \centering
  \includegraphics[width=\linewidth]{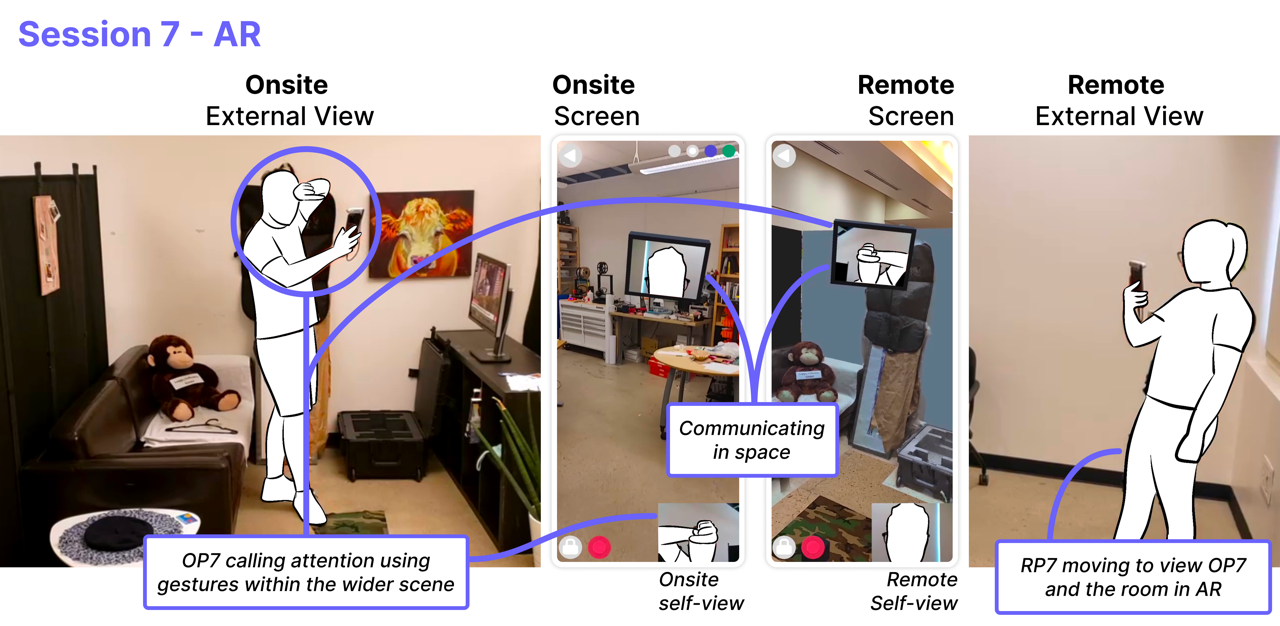}
  \caption{AR call interaction from Session 7. (Left) A view from OP7's space and phone screen, depicting how they are using gestures in their self-video to point to objects in the space while looking at RP7. (Right) A view from RP7's phone screen and space, showing that they can view both the 3D model of the scene, and the video representation of OP7 pointing at an object.}
  \label{fig:ar-vignette-1}
  \Description{This figure consists of four images (with people similarly sketched over and anonymized) - (Left) An external view of OP7 in the Art Heist scene, holding their hand in front of their face and pointing at an object so that it is captured in their video feed. (Center-left) A screenshot from OP7's phone, showing how in AR mode they are able to see a floating, moving video of RP7. (Center-right) A similar screenshot from RP7's phone, showing they can view OP7's video (and them pointing) on the backdrop of the 3D model of the shared room. (Right) RP7 using their phone to view the AR call.}
\end{figure}

\paragraph{Session 7 - AR}
In Session 7 (\autoref{fig:ar-vignette-1}), both OP7 and RP7 made extensive use of the ability to interact with spatial representations of each other. OP7 would seek out RP7's spatial video, confirm that they were looking at each other, and point at objects in the room through their self-video. This enabled RP7 to connect OP7's gestures with the wider 3D scene available to them. Both OP7 and RP7 paid attention to where they were located in space, and used their own ``embodied'' representations to call attention to certain objects, or navigate around each other to avoid overlap. Even during more independent exploration of the scene (which was only possible because of RP7's direct access to the 3D scene in AR), both participants kept up the work of communicating with each other and sharing verbal cues of context.

\paragraph{Session 5 -  AR}
\begin{figure}[ht]
  \centering
  \includegraphics[width=\linewidth]{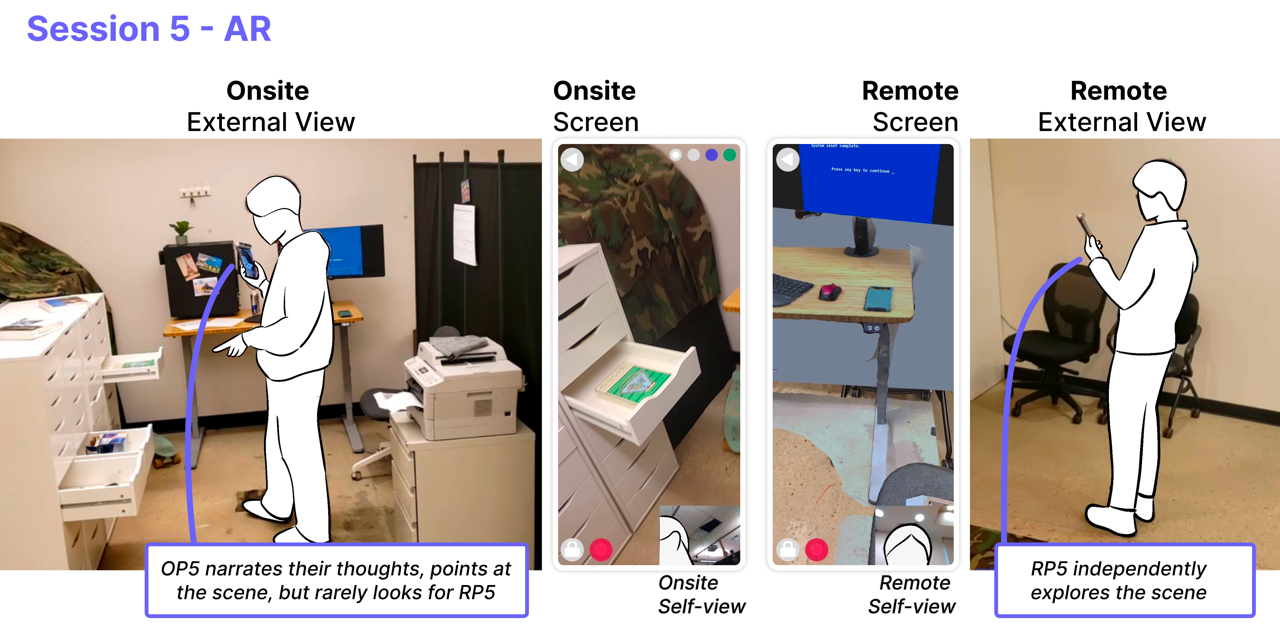}
  \caption{AR call interaction from Session 5. (Left) A view from OP5's space and screen, showing how they hold their phone to the side while pointing at objects for themselves during the exploration phase. (Right) A view from RP5's phone screen and space, as they independently explore a different part of the scene. While both OP5 and RP5 continue to narrate their thoughts, there is lesser face-to-face interaction in space.}
  \label{fig:ar-vignette-2}
  \Description{This figure consists of four images (with people similarly sketched over and anonymized) - (Left) OP5 is pointing at the shelves in the tech hack scene with their left hand, but they are holding their phone with their right hand in such a way that it is not capturing either their left hand or the object they are viewing. (Center left)  A screenshot from OP5's phone, showing the unframed video. (Center right) A screenshot from RP5's phone, where they are looking at the phone case on the standing desk. (Right) And external view of RP5 using the phone to view the shared room in AR.}
\end{figure}

Consider Session 5 (\autoref{fig:ar-vignette-2})---here, both OP5 and RP5 made use of the independence offered by AR to (1) view the real scene without the additional responsibility of framing the camera view (OP5), and (2) look at objects of interest without needing to wait or direct their collaborator (RP5). Both collaborators occasionally engaged in a more semantic discussion of the clues and what they meant towards the hypothesis, but they generally talked more about what was immediately on their mind. 
The focus on the real space meant that OP5 was not as conscious of their own representation in the shared space, and consequently, both participants did not engage in much face-to-face interaction. 
This led to small communicative breakdowns at times, where participants assumed they were talking about the same thing while observing different objects. 

\noindent
Across these four examples, we see hints of the importance of a shared sense of responsibility towards collaboration, how communicative work of different kinds is key across both mediums, and the potential benefits that spatial interaction can provide if navigated well.

\section{Findings}
\label{sec:findings}

We organize the discussion of findings and generated themes into three main sections. First, we highlight the key difference and benefit of AR calls in comparison to video---the ability to have increased spatial interaction within the shared environment and with each other. We then discuss how the medium of communication distributes agency and responsibility across onsite and remote participants. Finally, we highlight how the language of collaboration and the content that needs to be communicated for effective group work changes across both mediums. \autoref{tab:summary-table} provides an overview of these findings.

\begin{table}[ht]
    \centering
    \caption{A summary of findings organized by high-level themes and differences across AR calls and video calls. The three research questions are included at the bottom, and referenced alongside the themes.}
    \small
    \begin{tabular}{>{\raggedright\arraybackslash}p{9em} | p{0.1em} p{14em} | p{0.1em} p{14em}}
        \toprule
         & \multicolumn{2}{c|}{\textbf{AR calls}} & \multicolumn{2}{c}{\textbf{Video calls}} \\
         \midrule
         
         \footnotesize \textbf{The value of spatial interaction} \textit{(RQ1, RQ2)} & 
         \textcolor{ForestGreen}{\footnotesize \checkmark} & \footnotesize Like onsite participants, remote participants can have a more independent and embodied experience, spatially perceiving and exploring the shared environment & 
         \textcolor{BrickRed}{\footnotesize ?} & \footnotesize Remote participants can only view a section of the space at a time, and rely on the onsite participant's camera work and descriptions to build a spatial understanding of the room \\

         & \textcolor{ForestGreen}{\footnotesize \checkmark} & \footnotesize Both participants can observe and interact with each other in space, by moving around or talking face-to-face, similar to how people collaborate in real life &
         \textcolor{BrickRed}{\footnotesize ?} & \footnotesize While participants can interact with each other's video on-screen, focus might be more on the real environment or video, leading to unframed self-views \\

         \midrule

         \footnotesize \textbf{The medium distributes agency and responsibility} \textit{(RQ3)} & \multicolumn{2}{p{15.2em}|}{\footnotesize Greater agency for both onsite and remote participants, but shared responsibility to maintain collaboration} & 
         \multicolumn{2}{p{15.2em}}{\footnotesize Onsite participants have full agency and greater responsibility to maintain collaboration} \\
         & \textcolor{ForestGreen}{\footnotesize \checkmark} & \footnotesize Increased high-level sense of collaboration &
         \textcolor{ForestGreen}{\footnotesize \checkmark} & \footnotesize Helps participants be ``on the same page'' \\
         & \textcolor{BrickRed}{\footnotesize ?} & \footnotesize Can lead to decoupled experiences of the space &
         \textcolor{BrickRed}{\footnotesize ?} & \footnotesize Can push remote participants into a secondary role \\

         \midrule

         \footnotesize \textbf{The medium changes the vocabulary and content of communication} \textit{(RQ3)} & \multicolumn{2}{p{15.2em}|}{\footnotesize The shared room acts as a conversational resource, with participants using their words and bodies to maintain awareness of each other and coordinate joint action} & 
         \multicolumn{2}{p{15.2em}}{\footnotesize The shared video feed allows onsite participants to use deictic gestures, and both participants can use fragments of words and visuals to convey meaning} \\
         & \textcolor{ForestGreen}{\footnotesize \checkmark} & \footnotesize Supports embodied interaction, and participants can use semantic communication to connect to the space &
         \textcolor{ForestGreen}{\footnotesize \checkmark} & \footnotesize Helps establish a common ground for discussion \\
         & \textcolor{BrickRed}{\footnotesize ?} & \footnotesize Can cause breakdowns in communication due to a lack of awareness of each other's location and focus &
         \textcolor{BrickRed}{\footnotesize ?} & \footnotesize Communicating both semantic and spatial information can be challenging, particularly when the shared video is not intentionally framed \\

         \midrule 

        \multicolumn{5}{p{41em}}{\footnotesize \textit{(RQ1) How do people perceive and experience shared spaces when mediated by video and AR calls?}} \\
        \multicolumn{5}{p{41em}}{\footnotesize \textit{(RQ2) How do people perceive and interact with representations of each other in the real and remotely shared space?}} \\
        \multicolumn{5}{p{41em}}{\footnotesize \textit{(RQ3) How does a common understanding of the shared space emerge across local and remote collaborators?}} \\
         
         \bottomrule
    \end{tabular}
    \label{tab:summary-table}
\end{table}

\subsection{The value of spatial interaction with AR calls}

\subsubsection{Sharing spaces, together}
For remote participants, the most immediate benefit of the AR call was the presence of a 3D representation of the shared space that they could independently explore. In turn, onsite participants were freed from the responsibility of controlling the shared video viewpoint and could explore more independently as well. Over the course of the task, some participants began to build an understanding of the ability to view and interact with each other.

Both onsite and remote participants went through different stages of understanding how they inhabit the ``same'' space. Participants began by noticing their collaborator's video feed, and as they watched the video move in space, they developed an understanding of how it represents the location and focus of their partner in the room. In many sessions, participants positively remarked about knowing what their collaborator was looking at through their representation. These interactions also faced a few issues, most notably when participants would see the back of their collaborator rather than the front (which is where the live video feed was displayed) while looking at the same object.
We observed many cases where they playfully asked their collaborators to move out of the way, or themselves moved around to see the room more clearly.

Eventually, some participants began to make use of their own representations to direct attention to objects for discussion. The most striking example of this occurred in Session 7, where OP7 consciously moved themselves and pointed at objects through their self-video feed so that RP7 could gain further context about different points of interest. During certain moments in conversation when participants were called by their collaborators or felt the need to look for them, they sometimes did so by first looking at the last spatial location they remembered seeing their collaborator. While there were many instances early in the sessions where participants would pass through each other's representations, or stand very close to each other and block their collaborator's view, in the cases where participants became more aware of their own representations, they began to maneuver around and maintain a certain distance between each other, much like how one does in real life.

Thus, across sessions, we see how a growing awareness of the space and of each other's presence in the space was differentially navigated by each pair of participants. While some participants saw these ``blobs'' or ``rectangles'' as brief distractions from the goal of looking at the space for clues, others greatly valued the ability to look at each other while coordinating joint action.

\subsubsection{How bodies matter}
While the nature of the task---exploring a small room for clues pertaining to a mystery---is bound to lend itself to spatial movement in the scene, we see many clear examples of the importance of making sense of the space with respect to, and using, one's own body. The onsite participants, by virtue of being in the real space, were able to use their hands to touch objects while talking about them, crouch and kneel to gain a better view of objects, feel objects to get a sense of their state (such as tapping a soft-drink can to check its weight), and even use their bodies to make guesses about the mystery. Of particular note was OP3's reasoning about the height and handedness of the fictional suspect, given how the arrangement of objects in space was similar to their own preferences in real life. This form of embodied reasoning extended across different phases of the task, from building a coherent understanding of the scene, to thinking of hypotheses with the room as a resource. Across video and AR calls, we see many onsite participants rapidly move between looking at different objects while building theories, and sequentially step through the room as they recount those theories to themselves and their remote collaborators. 

When remote participants had spatial access to the scene during AR calls, they were able to engage in many of the same operations. For instance, RP4 walked around the space, looking at relevant objects while remembering the questions they needed to answer. RP10 took a step back and panned across the scene while summarizing their thoughts, and RP9 actively used their phone to tell a story through space, knowing that OP9 was able to see their representation move while doing so. Remote participants made use of their agency to move around the scene, explore based on interest, and reconfigure themselves around objects to view them more clearly. 

Across participants and sessions, we observe trends and archetypes of such embodied reasoning. Some participants act as surveyors, choosing a set of fixed vantage points from which to build a picture of the space. Others systematically scan the environment, moving from overview to detail views and back again. Others still are more opportunistic, and use their ability to freely move around to do so in more spontaneous ways, based on the theories they may be thinking of at the time. All the while, participants are able to watch and seek out their collaborators to ask questions and discuss their hypotheses. A crucial benefit of AR calls is that both onsite and remote participants have the ability to make use of their bodies and explore the space in the way that best suits them. Thus, unlike the one-directional dependence of video calls, AR calls help bring the remote collaborative experience closer to what it would be like in real life.

\subsubsection{Quantitative hints}
For a sense of how participants rated their level of spatial presence across the various conditions, and how they fared with the spatial mapping post-task exercise (which was rated by the research team), we provide an overview of key statistics in \autoref{fig:quant-figure} and \autoref{tab:quant-table} (in the appendix). Through these figures, we see initial hints of how AR calls are rated as providing both onsite and remote participants with a greater sense of spatial presence, and help remote participants improve their recall of the space. 
In keeping with the spirit of Comparative Structured Observation, our focus in this paper is on a qualitative understanding of participant experience. We do not conduct an extensive quantitative analysis testing for significance of differences---the sample size makes such claims difficult in any case. Instead, these initial quantitative findings reinforce the reported and observed benefits of spatial interaction, and help add to the backdrop of our discussion of subsequent themes.

\begin{figure}[ht]
  \centering
  \includegraphics[width=\linewidth]{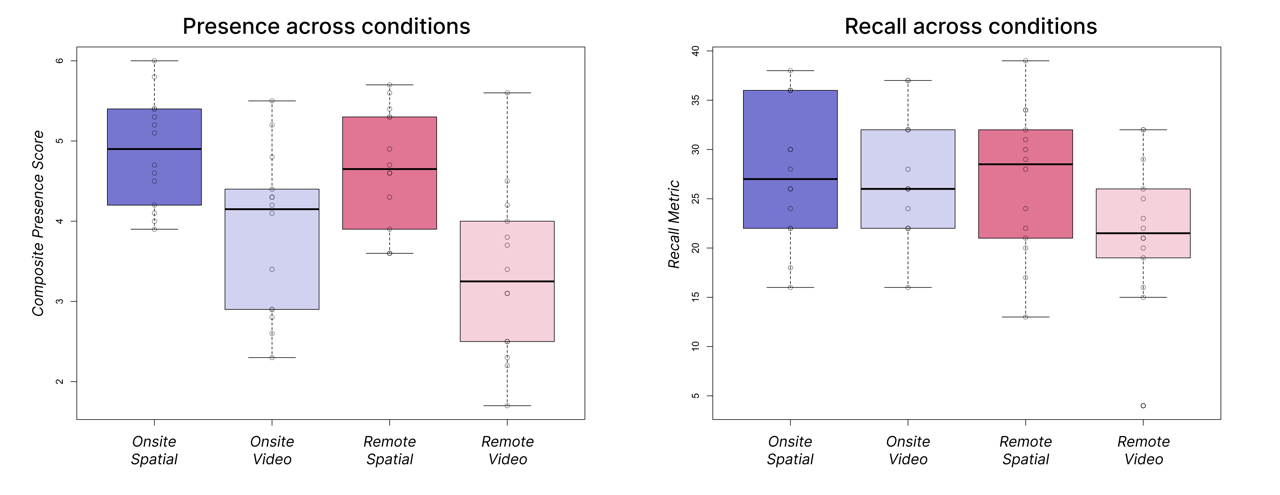}
  \caption{Spatial presence and recall scores across the four possible states of Onsite Spatial, Onsite Video, Remote Spatial, and Remote Video.}
  \label{fig:quant-figure}
  \Description{On the left of this figure is a box plot depicting changes in presence scores across conditions, and on the right is a box plot depicting the change in recall scores across conditions. For each box, we also see the mean, standard deviation, and individual data points marked. Going from onsite spatial to remote video, the color of the boxes moves from purple to light red. In the presence plot, we see that Onsite Spatial and Remote Spatial are quite higher than the video conditions. Onsite spatial is slightly higher than remote spatial, while remote video is slightly lower than onsite video. For the recall plot, onsite spatial is the highest, onsite video and remote spatial are comparable, while remote video is the lowest.}
\end{figure}

\subsection{The medium distributes agency and responsibility}

\subsubsection{Lopsided responsibilities in video calls}

Due to the inherent asymmetry of information access in video calls, the onsite participant was responsible for the task of sharing information from the physical space with the remote participant, in addition to collaborating with them to look for clues and develop hypotheses. The first of these responsibilities weighed on some participants, with OP14 mentioning they felt like they were their collaborator's \textit{``eyes and ears''}.

\noindent
As a result, onsite participants took on the role of leading the collaboration, with some remote participants largely being along for the ride.
In some cases, onsite participants tried to give their remote collaborators more of a role in influencing the course of the exploration, but they were ultimately in control of the extent to which this freedom was provided.
The asymmetry of access and control caused remote participants to take a step back when the onsite collaborators did not keep remote needs in mind. The remote role was seen as being more focused on the higher-level task of solving the mystery and communicating with the onsite participant to clarify details relevant to that goal.
Participants remarked how their task was reduced to trying to \textit{``stare at the phone better''} (RP7, Video), and how the scope of their ability to explore was limited to \textit{``the peripheries''} (RP13, Video) of what their onsite collaborators were sharing with them.

This ceding of control was in some cases accompanied by the assumption that the onsite participant was (1) capturing all the important details while (2) intentionally framing the camera view to show the remote participant what they needed to see. This, however, was not always the case, as onsite participants mentioned forgetting about their remote collaborators during periods of silence when focused on solving the mystery.
Some onsite participants mentioned how they were not very collaborative by nature, and liked having their remote collaborators be a lot more dependent on them.
Thus, the interests of the onsite participant often took precedence, with less focus on the communicative work required to meaningfully collaborate on the task.

In video calls, the onsite participant is the only one who has a spatial perspective of the scene. Their remote collaborator can only view a slice of this scene on their phone. Some participants found that the separation of roles as a result of this constraint, and the need to converse more, helped keep discussions about their hypotheses on track.
With video, there is far less communication needed to establish a collaborator's point of view, but far more to change and direct that point of view due to the asymmetry of control. In some cases, the increased sense of responsibility in video calls caused some onsite participants to consciously make a plan for effective exploration: \textit{``I did feel like I had to help you get a sense of the space and like, I even had my whole strategy''} (OP7, Video).
Thus, video calls, by nature of the medium, support (and force) a singular shared perspective of the scene---one that while conducive to focused discussions, is dictated by the performance and interests of the onsite collaborator.

\subsubsection{Shifting roles with the independence of AR calls}
The ability to independently explore the space provided remote participants with a lot more agency in the AR call. They were able to walk around, look at objects, verify details mentioned by their collaborators, and work in parallel to look at different parts of the room.
Participants remarked how not having to wait for their onsite collaborators to show them information, and being able to \textit{``wander off''} (RP8, AR) on their own while being aware of their collaborator's location, helped them feel like their role in the task was more significant. This independence also affected the onsite participants' perception of the task, or as OP7 succinctly put it:
\textit{``It was helpful for me for her to have her own autonomy''}.

For remote participants, the ability to go \textit{``poking around''} (RP1, AR) gave them more agency, and helped build a stronger sense of space, resulting in mental maps and drawn maps that they perceived to be more clear.
In many cases, participants continued to communicate with each other to discuss ideas and provide spatial context for their hypotheses.
Some participants connected this independence of spatial perspective to an increased ability to form their own perspectives about the scene and the mystery, in contrast to video calls where remote participants have to rely on their onsite collaborator's experience, which is bound to color their own interpretation.
These different perspectives enable both participants to collaborate, to some extent, as equals, with RP5 noting how \textit{``both of us can take the lead and find our own clues''}.
However, at times, the independence provided by the AR call caused the collaborative task to be decomposed into two separate explorations,  
with remote participants being more focused on the space than the moving representation of their collaborator. 
Despite this decoupling, the increased sense of control perceived by remote participants (thereby allowing them to play more of an active role in the exploration) resulted in an interesting contrast between the enaction and perception of joint activity, with RP3 noting how by \textit{``both doing our own thing ... we're collaborating''}.

\subsection{The medium changes the shared vocabulary and the content of communication}

\subsubsection{The language of AR calls: sharing spatial context with words and bodies}

With the remote participants having independent access to the scene in AR calls, onsite collaborators no longer needed to engage in extensive visual communicative work, focusing instead on the physical space in front of them while using the phone to speak across the spaces. 
In some cases, participants capitalized on this mutual access and used the shared room as a conversational resource by employing more spatial references to objects and directions: \textit{``I already know that he knows what's going on in their room ... I can just tell him, oh, like, I'm looking at the drawer, like, let's look at them together''} (OP5, AR).
As the session progressed, some onsite participants realized that new forms of communicative work were required in AR to navigate potential breakdowns in communication. For instance, OP4 noted how \textit{``if I'm pointing the camera at something, I don't know if she's looking at the same thing ... I was talking about one book and she was talking about some other book at that point. So I didn't realize it until she said, the title of the book''} (OP4, AR).
With this awareness, participants put more effort into providing spatial context, looking for where their partners were in space, and being more conscious of their positioning when \textit{``pointing in that direction''} (OP7, AR).
Thus, both collaborators need to do the work of maintaining a shared sense of space and context in AR. 

During AR calls, we saw many instances of participants referring to ``that'' object that is present ``here'', often followed by requests to disambiguate where ``here'' even is. In response to this, we see examples of both onsite and remote participants doing the work of narrating their point of view, moving their bodies consciously to indicate focus, and looking for their collaborators in space to be aware of their point of view. Unlike video calls, where the shared pane of visual context is entirely controlled by the onsite participant, in AR we see how a shared model of the space is constructed over time. Both collaborators are able to move around in the scene while discussing specific objects and their hypotheses around them, and this leads to the development of a shared understanding where towards the end of sessions, even semantic discussions provide enough context for participants to know what their collaborator is talking about.

\subsubsection{The language of video calls: sharing every detail, with a shared pane of context}

A key struggle caused by the asymmetry of access in video calls is the need for more coordinative conversations, where remote participants have to ask their onsite collaborators to move a certain way, or return to particular objects: \textit{``it also calls for so much more communication as well. It's a lot more frustrating to work with someone in video because you're like, go here go there, no show me this''} (RP3, video).
This constant back and forth at times reduces the feeling of collaboration for remote participants.

The shared language of communication during video calls consists of verbal as well as visual channels---the framing of the environment view, and of each others' faces in the call are key aspects of the ``work'' that both onsite and remote participants need to actively engage in.
The presence of a shared visual field acts as a valuable conversational resource when used well. It enables conversations to consist of fragments of verbal and visual phrases, and at its best enables onsite and remote participants to be ``on the same page'' when discussing an object or the scene. 
When sharing video with their remote collaborators, some onsite participants would narrate additional details such as visual clues that might not be as clear through the video, the relative positioning of objects, as well as their own motion in space. 
Coupled with the real-world video feed, this verbal communicative work helped remote participants feel included and gain a better understanding of the scene.
We see many instances of onsite and remote participants using conversational references to objects on the screen---onsite participants would move their phone to point at an object, and ask their remote collaborators what they think about ``this''. Remote participants could ask questions based on the spatial relationships of objects currently displayed on the screen, such as requesting a more detailed view of a certain element on top of a piece of furniture.

By providing a narrative of their spatial movement and the prepositional relationships between objects in a scene, onsite participants help construct a shared image of the space that their remote collaborators can then use to direct movement later on. 
When accompanied with such additional channels of communicative work from the onsite participant, remote collaborators are able to gain a sense of intentionality in the onsite participant's actions, and can feel like their need to engage in additional confirmatory conversations (a key source of friction and disengagement) is reduced.
This additional emphasis on paying attention to and communicating every detail led some onsite participants to feel like they remembered more of what was discussed.
Some onsite participants consciously moved between detail and overview views, either in response to requests from their remote collaborators, or implicitly based on their perception of the needs of the conversation. 

However, the greater control in the onsite participant's hands meant that even attempts at providing richer descriptions came with editorial decisions based on their perception of what was most important in the moment: \textit{``I just like glanced and I just gave this summary, that would be quicker for me to say and for you to understand''} (OP4, Video). The shared ``page'' or ``pane'' of context on the screen is only useful when it is intentionally framed, and when onsite participants are also looking at it as their remote collaborators talk in relation to it. Not all onsite participants approach their role with this level of intentionality, leading to breakdowns when remote participants assume that their collaborators are looking at the same thing as they are, when in fact they are not.

Ultimately, the choice to engage with this communicative work rests largely on the onsite participant. When done well, as we have seen, onsite participants act on a perceived responsibility towards their remote collaborators, who in turn feel more included in the activity. When done poorly, the process of building a shared understanding of the space becomes fragmented, with the onsite participant dictating much of what takes place during the exploration.

\subsubsection{The role of gestures}
Deictic gestures are a key part of coordinating action in space, and onsite participants in particular made use of them across both video and AR calls. In video calls, they pointed at objects in such a way that their hand was visible in the shared video feed in order to direct their remote collaborator's attention, or clarify points of discussion. In many cases, the onsite participants pointed at objects in the space but outside the field of view of the phone camera, in ways that seemed to help their own reasoning in the moment rather than serve a communicative purpose. This extended to AR calls, where onsite participants would still point to objects and gesture at things in the room. In some cases, they soon realized that their collaborators cannot see their actions in the same way, yet the use of such gestures persisted. In contrast, the occurrence of gestures and pointing actions on the remote end was much lesser. During video calls, some participants tried pointing at objects on their screen, but during those moments, their onsite collaborators were often looking directly at objects in their local space, missing remote gestures on the screen. In AR calls, remote participants seem more focused on appropriately holding the phone to view the shared environment, which might be one reason for a reduction in hand gestures and pointing actions. We instead see cases in AR calls where participants use the movement of their representations to signal to their collaborators, as they develop an emergent sense of their own bodies in the shared space.

\subsubsection{Illusions of being on the same page}
The idea of ``being on the same page'' is an underlying goal across many of the collaborative sessions analyzed here. Given how video calls enable participants to have a shared ``page'' of context, and how AR calls require participants to ``read the room'' and build a shared understanding of space together, we find it useful to think of this form of collaboration through the metaphor of reading a book. Doing so helps us see the different ways in which the medium supports collaboration, and leads to illusions of being on the same page.

In video calls, the onsite participant simultaneously has full control on the shared video feed (the page), and complete freedom to physically look around in the local space (the story). We see numerous examples of onsite participants pointing their phone at a specific object or section of the scene, while continuing to look around and browse the space. At the same time, their remote collaborators only have access to what is being shared, and often assume that it is what the onsite participants are also focused on. Thus, while it may seem like both participants are looking at the same page, the onsite participant continues to read the rest of the story. Onsite participants can quickly scan or browse the broader scene to gain more details or answer initial questions. Remote participants are limited to verbally asking their onsite collaborators to turn the page, move, or return to specific objects, which is a much slower process. The act of asking questions for more context can make the difference in informational access more stark. In response to such questions, onsite participants sometimes move the phone very close to the relevant object such that it occupies the entire video feed---this might be useful to read smaller details, but more often than not, camera work trails the embodied experience of making sense of space. These phases increase the imbalance of contextual understanding across collaborators. Remote participants are forced to read a single word on the page, while their onsite partners continue to have full access to the entire story.

With AR calls, this imbalance is significantly reduced, as both onsite and remote participants have independent access to the same ``story''. Onsite participants can continue to explore the scene as they would normally do, but remote participants have the independence to read at their own pace, move between detail
and overview views as needed, and answer questions they might have of their own volition. However, this independence also comes with an asymmetry---remote participants can only view the scene through their phone, while onsite participants can continue to look around the real space. As a result, onsite participants in certain sessions are able to simultaneously engage in two operations: using the phone to keep track of their remote collaborator, while looking around the scene to pick up on details and form ideas.

Another challenge is the need to maintain awareness of each others' spatial context (where one is in the story, or how much they have read). In some cases, we observe participants in different locations in the room who think they are talking about the same object when they are not, leading to peculiar breakdowns that are sometimes never resolved within the session. Thus, these illusions of being on the same page take on different forms across video and AR, the former caused by an asymmetry of control and access, and the latter by the work needed to maintain shared context.

\subsection{Looking back}

To conclude this section, we provide a brief summary of findings in relation to our initial research questions. \autoref{tab:summary-table} contains a tabular version of this summary.

\subsubsection{How do people perceive and experience shared spaces?}

In video calls, onsite participants have full access to the physical space and complete control over their remote collaborator's view. As a result, remote participants' spatial perception is dependent on how well their onsite collaborator frames the camera view, provides spatial and semantic descriptions, and shares control over the exploration. The onsite participant is the only one ``in'' the space, with the remote participant being largely along for the ride.

With AR calls, remote participants gain much more independence to view and make sense of the shared space. They are able to take control of their perspective, and use their bodies to perceive and reason about space in much the same way as onsite participants can. In turn, onsite participants are freed from some of the responsibility of leading the call, and can explore the space in ways more directly dictated by their interests and by natural conversation.

\subsubsection{How do people perceive and interact with representations of each other?}

When collaborators are present as small videos fixed to the bottom of the phone screen during video calls, we see an expected range of behaviors. While some participants appreciated the constant, responsive view of their collaborators, we notice how onsite participants are more focused on the real-world scene than on their remote partner's face, which causes lesser face-to-face interaction in general.

In contrast, as participants grew more familiar with the affordances of the AR call, they began to interact with each other in ways quite similar to in-person communication. Both onsite and remote collaborators would keep track of each others' location and focus, turn to discuss their hypotheses face-to-face, remember and maintain their relative spatial positioning, and use their bodies to seek and direct attention while exploring the space.

\subsubsection{How does a common understanding of a shared space emerge?}

The medium of communication distributes agency and shapes the shared vocabulary for collaboration. However, participants' perception of their responsibilities, and the nature of verbal, visual, and spatial communication during joint action, are key drivers of the development of a common understanding of shared space. In video calls, the shared ``page'' of visual content helps provide common ground during discussion. Onsite participants have a greater responsibility to guide the creation of a shared model of the space for them and their remote collaborators. With AR calls, both onsite and remote participants can use words and their bodies to communicate awareness and discuss ideas, while exploring the scene in their own styles and time. Taken together, AR calls help support collaboration that comes much closer to in-person spatial interaction than what video calls provide.

\section{Discussion}

\subsection{Studying the landscape of mobile spatial collaboration}
Mobile video calls have become a key fixture of the modern communication landscape, and as a result the dynamics of collaboration over mobile video have been extensively studied in real or realistic contexts over the years \cite{licoppe2009CollaborativeWorkProducing,brubaker2012FocusingSharedExperiences,jones2015MechanicsCameraWork}. Our findings from this study echo key insights from these prior works. In particular, we see many of the same mechanics and challenges of camera work persist in mobile video interaction today, as presented by \citet{jones2015MechanicsCameraWork}. Onsite participants used a similarly wide range of techniques to provide overview and detail views to their remote collaborators, and camera work continues to complement verbal communication as onsite and remote participants make use of fragments of words and visuals. Remote participants still need to \textit{``integrate multiple scenes over time, remembering what they saw in relation to what they are seeing now''}, and when particular views of the shared space are not available, requesting the onsite participant to provide those views remains \textit{``awkward and cumbersome''}. Despite a decade separating Jones et al.'s work and this paper, the authors' statement that \textit{``current designs of mobile video conferencing technologies are mainly the mobile equivalent of their desktop counterparts''} still applies today, with the only new feature of note being the ability to view the video feed of peoples' faces and their surroundings simultaneously (when using specific mobile video platforms).

The idea of using AR to provide more spatial forms of interaction is an attempt to move past ``mobile desktop'' implementations. However, most studies of this medium are tied to early-stage prototypes tested in simplified usage contexts, where the contribution focuses on the technical novelty of the prototype. Prior projects such as ColabAR \cite{villanueva2022ColabARToolkitRemotea} and ARCritique \cite{li2022ARCritiqueSupportingRemote} have demonstrated the value of mobile AR in supporting object-focused discussions in university laboratories and design studio courses. In these and many other mobile AR projects \cite{muller2017RemoteCollaborationMixed}, the collaborators are represented as simple 3D models in space. Mobileportation \cite{young2020MobileportationNomadicTelepresence} and DualStream \cite{vanukuru2023DualStreamSpatiallySharing} go beyond this by developing novel prototypes for remote collaborators to share real-time representations of their selves and surroundings. Both projects presented promising initial findings from pilot studies where individual participants used the prototypes to interact with members of the research team. 

In this paper, we sought to look beyond the prototype. By focusing on studying the nature of collaboration between two individuals who are familiar with each other, and conducting a systematic comparison between mobile AR and mobile video, our work is among the first few projects to provide an in-depth understanding of mobile AR collaboration in realistic (if still experimental) conditions. Going beyond the more obvious benefits of AR interaction---the independent spatial perception and exploration of remote spaces---we highlight how AR calls help support dynamics of collaboration that come quite close to in-person interaction. Our findings paint a nuanced picture about the relative benefits and limitations of both video and AR calls, which we hope sets the stage for future design and evaluation work that helps realize the potential of mobile spatial collaboration in the context of real-world use. 

\subsection{From \textit{showing} to \textit{sharing} space}
The model of collaboration with mobile video revolves around ``showing'' objects and spaces to people who are not physically present. But while visual information of a scene may be primarily transmitted in one direction (from local to remote), knowledge about it might not be. Calling a relative to show them the layout of your new apartment is very different from calling a colleague to ask for help when looking for something in their office, for example. The needs of agency and the distribution of responsibility vary greatly across these contexts. Prior studies of mobile video have tried to capture this range of contexts in their choice of experimental tasks \cite{jones2015MechanicsCameraWork}, but in all cases, mobile video imposes a fixed set of constraints regarding who has agency and control.
With AR calls, this agency and control is distributed equally across onsite and remote collaborators, enabling them to ``share'' a space together. 
An apartment tour \cite{young2020MobileportationNomadicTelepresence} or design critique session \cite{li2022ARCritiqueSupportingRemote} might be best conducted using AR calls where both parties have independent access to and opinions about the scene, but more focused search or assistance tasks might still benefit from the specific mechanics of video calls. 

Returning to the idea of ``sharing'' space, AR calls help bring back aspects of ``workspace awareness'' \cite{gutwin_descriptive_2002} to mobile video communication. As Gutwin and Greenberg put it, \textit{``the first information source is the other person’s body in the workspace''}, and both onsite and remote bodies begin to matter more when communicating in AR. By supporting interactions that closely approximate in-person collaboration, people can better manage coupling during tasks, coordinate actions, and convey intentions through gestures and mutual awareness. 
Our theme titled ``how bodies matter'' is a reference to a foundational paper on embodied forms of interaction design \cite{klemmer2006HowBodiesMattera}. With mobile AR, we see the themes of visibility, performance, and thinking through doing become more apparent as people use their bodies to explore, communicate, and make sense of space.

\subsection{Towards design}
A key goal of this project was to inform the design of future tools for mobile spatial collaboration, based on insights from a controlled evaluation of both video calls and AR calls. To that end, our study provides strong evidence of the benefits of AR calls, while also highlighting the aspects of video calls that make them very effective in certain contexts.

\noindent
In terms of specific features and interactions that can be improved, our participants shared many ideas that are worth exploring. For some, the representation of their remote collaborators could have used more detail. OP11 remarked that they would have liked to see a representation that \textit{``more closely resembles a human''}, potentially supporting more direct deictic gestures.
Building upon prior work, we imagine mobile AR setups where screen-based interactions can result in digital deictic gestures (like laser pointers into the shared space, similar to \cite{vanukuru2023DualStreamSpatiallySharing}), or where the movement of the phone in space could be used to animate articulated 3D avatars \cite{ahuja2021PoseontheGoApproximatingUser}, while retaining the ``real'' aspects of participants' self-video feeds.
The nature of the 3D scan reminded some participants of more conventional 3D software and model inspection tools, leading to ideas for more degrees of freedom in exploring the remote scene, such as scaling or zooming into the model (OP4).

Some participants made comparisons to how existing groupware tools support workspace awareness such as cursor indicators in collaborative document editors (OP13), and wondered if similar approaches could be applied to mobile AR.
Prior work in AR/VR collaboration tools has explored how ``halo'' indicators \cite{baudisch2003HaloTechniqueVisualizing}, mini-maps, or other edge-of-screen notifications \cite{gruenefeld2018HaloWedgeVisualizing} can help provide people with more of a sense of where their collaborators are when outside their immediate field of view, and we can see similar interaction techniques being applied to mobile AR as well.

For participants who had prior experience with headset-based AR and VR, one common point of discussion was picturing how AR calls would work with headsets. While some participants pointed to the inevitability of headset-based interaction, which prior work seems to indicate  might be preferable, headsets are still far from widespread adoption. Very few participants had VR/AR headsets of their own, and were excited to see similar spatial features being implemented using mobile devices.
In addition to benefits offered by the wider access of mobile phones, one participant (RP7) mentioned that the AR call caused them to experience less motion sickness than they normally get from head-worn devices, and in some ways mobile AR felt perceptually better than the video call as well. Even in a near-future where headsets are more commonplace, there are many reasons why people might not be able to use them, and mobile AR experiences can help widen the range of people who have access to spatial forms of collaboration and interaction.

\begin{figure}[ht]
  \centering
  \includegraphics[width=\linewidth]{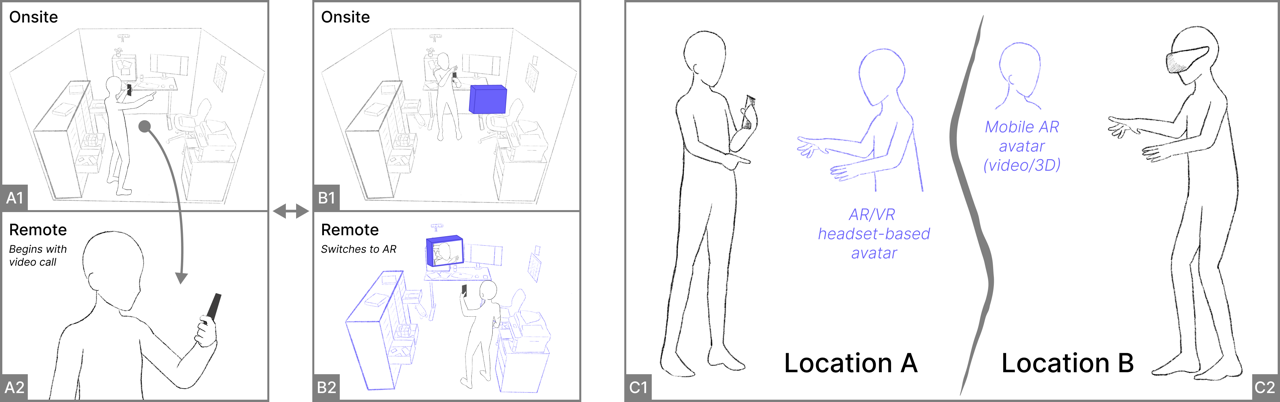}
  \caption{Future visions of mobile spatial collaboration. (Left) An onsite user could begin sharing a view into their space (A1) while their remote collaborators view a video (A2). (Center) Once the application builds a 3D model of the shared scene, both collaborators (B1, B2) can switch to AR mode, interacting with the room and each other in space. (Right) Mobile AR can be used to interact with people wearing headsets, where a moving mobile video or 3D representation can serve as the mobile ``avatar'' that a headset-wearing collaborator views.}
  \label{fig:future-work}
  \Description{This is a composite figure consisting of three columns of content. (Left top) A perspective sketch of an onsite person using a phone for a video call while pointing at an object in their space. (Left bottom) A sketch of a remote person viewing the scene on their phone. (Center top) The same perspective sketch, but now the onsite person can see a floating video of their remote collaborator. (Center bottom) A perspective sketch showing how the remote person can view both their onsite collaborator and a 3D scan of the shared space in AR. (Right) A sketch showing how a mobile AR user can speak to a 3D avatar of a headset-wearing user, while the person with the headset can see a mobile-generated 3D representation, as a vision for the future of cross-device communication.}
\end{figure}

\noindent
Ultimately, we see mobile AR as another point in the broader continuum of mobile communication technologies. Our study shows that both mobile video and mobile AR have their place in a future suite of spatial collaboration tools. The natural approach, as pointed out by OP7, would be to blend the two modalities, and offer users the choice to switch between them spontaneously: \textit{``even a small hybrid of, those could be really interesting where it's just like here. Let me actually share with you what I'm looking at currently (while in AR). And then get the video portion. But then you can elect to go in and out of that''} (OP7).

Even within the context of a short spatial investigation task such as ours, we see how participants move in and out of phases of exploration and discussion, with differential use of spatial elements. As we move towards future design, we envision mobile AR becoming more seamlessly integrated with mobile video communication (\autoref{fig:future-work}, Left). For example, during the first few seconds of a video call where someone is sharing a view into their local space, the application could create a 3D model of the space in the background. Once the model is processed, the collaborators could move in and out of AR mode based on the needs of the conversation, and even blend between content shared from multiple locations at once. These interaction techniques could extend to more cross-device configurations of collaboration, where people can use whatever devices they have access to---mobile phones, tablets, headsets---to meaningfully collaborate with each other within and across space (\autoref{fig:future-work}, Right).

\subsection{Limitations and future work}

We approached the design, execution, and evaluation of this study keeping in mind the guidelines and best practices of Comparative Structured Observation \cite{mackay2025ComparativeStructuredObservation}. However, just as with most controlled experiments of interactive prototypes, our choices come with their own limitations. Chief of these pertains to the task and its bounds. We chose a spatial investigation task to ensure participants would be sufficiently challenged and motivated to explore the scene, and chose not to ``force'' or ``simulate'' the need for collaboration by offering differential access to task-relevant information. 
Despite this, all participants collaborated with each other extensively and in much more natural ways than we could have engineered. The loose upper time limit for each task led to some sessions being cut short before the participants could fully discuss their hypotheses. Our participant pool also largely consisted of university students with above-average experience with AR and VR technologies. These limitations can only be meaningfully addressed through real-world studies of use, and we believe that our findings provide strong evidence to justify and motivate such studies in the near future.

This paper focuses on a rich, qualitative analysis of user experiences first, with support from brief analyses of some quantitative data. In the process of conducting the experiment, collecting the data, and processing it for more meaningful use, we have created a wide range of data sources (motion log visualizations, participant focus graphs, hand-drawn maps) that could form the basis for a more spatially-oriented analysis of movement, coordination, and joint focus. In the future, we plan to complement the analyses in this paper with a more exploratory quantitative analysis of our dataset, to see if we can uncover further patterns of communicative work that are not directly expressed through the video recordings and transcripts. While we were only able to present initial findings related to participant recall from their hand-drawn maps of the space, one direction of particular interest is to further analyze these maps to understand how the medium of collaboration influenced the construction of spatial mental models across participants.

Moving forward, we plan to develop prototypes that address some of the interaction challenges mentioned by participants, and create tools that allow for a more smooth transition between video and AR collaboration. Just like how real-world studies of mobile video use have provided deeper insights \cite{licoppe2009CollaborativeWorkProducing, ohara2006EverydayPracticesMobile}, we hope to conduct more longitudinal evaluations of the ideas presented here. Instead of testing for a wide range of simulated tasks in laboratory settings, such a real-world evaluation would help identify the contexts in which AR and video are likely to be most useful, help users grow familiar with both modalities, and provide richer insights about the potential benefits of spatial tools for mobile-based collaboration.

\section{Conclusion}
Mobile video calls are among the most widely-used forms of ``spatial'' communication over distance, but lack much of the spatial perception and interaction possibilities of in-person collaboration. Recent work on mobile AR systems for remote collaboration shows promise in addressing this gap. In this paper, we conduct a comparative structured observation study to understand how pairs of participants experience and reflect on collaboration across mobile video and mobile AR. We find that the medium influences how collaborators negotiate their roles and responsibilities, and changes the nature of communicative work required for coordination. Mobile video helps collaborators to be ``on the same page'' and use the shared video as a conversational resource, while mobile AR gives both onsite and remote collaborators independent access to the same spatial scene, resulting in more natural forms of conversation.
While onsite participants make extensive use of their bodies to reason about space and convey meaning to their remote collaborators, mobile AR helps remote participants engage in similar forms of embodied interaction independently. Taken together, we discuss how each modality influences the way we interact with each other and build a shared understanding of collaborative tasks, and offer insights and implications that can help shape the design of future systems for mobile spatial collaboration.

\begin{acks}
We would like to thank all of our study participants for their time and feedback. We thank members of the ACME Lab for their assistance during the user study, Sandra Bae and Takanori Fujiwara for helpful discussions about data analysis, and the anonymous reviewers for many constructive comments that helped shape the final version of this paper. This project was supported by a grant from Ericsson Research.
\end{acks}


\begin{thebibliography}{50}


\ifx \showCODEN    \undefined \def \showCODEN     #1{\unskip}     \fi
\ifx \showISBNx    \undefined \def \showISBNx     #1{\unskip}     \fi
\ifx \showISBNxiii \undefined \def \showISBNxiii  #1{\unskip}     \fi
\ifx \showISSN     \undefined \def \showISSN      #1{\unskip}     \fi
\ifx \showLCCN     \undefined \def \showLCCN      #1{\unskip}     \fi
\ifx \shownote     \undefined \def \shownote      #1{#1}          \fi
\ifx \showarticletitle \undefined \def \showarticletitle #1{#1}   \fi
\ifx \showURL      \undefined \def \showURL       {\relax}        \fi
\providecommand\bibfield[2]{#2}
\providecommand\bibinfo[2]{#2}
\providecommand\natexlab[1]{#1}
\providecommand\showeprint[2][]{arXiv:#2}

\bibitem[Ahuja et~al\mbox{.}(2021)]%
        {ahuja2021PoseontheGoApproximatingUser}
\bibfield{author}{\bibinfo{person}{Karan Ahuja}, \bibinfo{person}{Sven Mayer}, \bibinfo{person}{Mayank Goel}, {and} \bibinfo{person}{Chris Harrison}.} \bibinfo{year}{2021}\natexlab{}.
\newblock \showarticletitle{Pose-on-the-{Go}: {Approximating} {User} {Pose} with {Smartphone} {Sensor} {Fusion} and {Inverse} {Kinematics}}. In \bibinfo{booktitle}{\emph{Proceedings of the 2021 {CHI} {Conference} on {Human} {Factors} in {Computing} {Systems}}} \emph{(\bibinfo{series}{{CHI} '21})}. \bibinfo{publisher}{Association for Computing Machinery}, \bibinfo{address}{New York, NY, USA}, \bibinfo{pages}{1--12}.
\newblock
\showISBNx{978-1-4503-8096-6}
\href{https://doi.org/10.1145/3411764.3445582}{doi:\nolinkurl{10.1145/3411764.3445582}}


\bibitem[Arth et~al\mbox{.}(2015)]%
        {arth2015HistoryMobileAugmented}
\bibfield{author}{\bibinfo{person}{Clemens Arth}, \bibinfo{person}{Raphael Grasset}, \bibinfo{person}{Lukas Gruber}, \bibinfo{person}{Tobias Langlotz}, \bibinfo{person}{Alessandro Mulloni}, {and} \bibinfo{person}{Daniel Wagner}.} \bibinfo{year}{2015}\natexlab{}.
\newblock \bibinfo{title}{The {History} of {Mobile} {Augmented} {Reality}}.
\newblock
\href{https://doi.org/10.48550/arXiv.1505.01319}{doi:\nolinkurl{10.48550/arXiv.1505.01319}}


\bibitem[Baudisch and Rosenholtz(2003)]%
        {baudisch2003HaloTechniqueVisualizing}
\bibfield{author}{\bibinfo{person}{Patrick Baudisch} {and} \bibinfo{person}{Ruth Rosenholtz}.} \bibinfo{year}{2003}\natexlab{}.
\newblock \showarticletitle{Halo: a technique for visualizing off-screen objects}. In \bibinfo{booktitle}{\emph{Proceedings of the {SIGCHI} {Conference} on {Human} {Factors} in {Computing} {Systems}}} \emph{(\bibinfo{series}{{CHI} '03})}. \bibinfo{publisher}{Association for Computing Machinery}, \bibinfo{address}{New York, NY, USA}, \bibinfo{pages}{481--488}.
\newblock
\showISBNx{978-1-58113-630-2}
\href{https://doi.org/10.1145/642611.642695}{doi:\nolinkurl{10.1145/642611.642695}}


\bibitem[Benford et~al\mbox{.}(1998)]%
        {benford1998UnderstandingConstructingShared}
\bibfield{author}{\bibinfo{person}{Steve Benford}, \bibinfo{person}{Chris Greenhalgh}, \bibinfo{person}{Gail Reynard}, \bibinfo{person}{Chris Brown}, {and} \bibinfo{person}{Boriana Koleva}.} \bibinfo{year}{1998}\natexlab{}.
\newblock \showarticletitle{Understanding and constructing shared spaces with mixed-reality boundaries}.
\newblock \bibinfo{journal}{\emph{ACM Transactions on Computer-Human Interaction}} \bibinfo{volume}{5}, \bibinfo{number}{3} (\bibinfo{date}{Sept.} \bibinfo{year}{1998}), \bibinfo{pages}{185--223}.
\newblock
\showISSN{1073-0516}
\href{https://doi.org/10.1145/292834.292836}{doi:\nolinkurl{10.1145/292834.292836}}


\bibitem[Billinghurst and Kato(2002)]%
        {billinghurst2002CollaborativeAugmentedReality}
\bibfield{author}{\bibinfo{person}{Mark Billinghurst} {and} \bibinfo{person}{Hirokazu Kato}.} \bibinfo{year}{2002}\natexlab{}.
\newblock \showarticletitle{Collaborative augmented reality}.
\newblock \bibinfo{journal}{\emph{Commun. ACM}} \bibinfo{volume}{45}, \bibinfo{number}{7} (\bibinfo{date}{July} \bibinfo{year}{2002}), \bibinfo{pages}{64--70}.
\newblock
\showISSN{0001-0782}
\href{https://doi.org/10.1145/514236.514265}{doi:\nolinkurl{10.1145/514236.514265}}


\bibitem[Biocca and Harms(2002)]%
        {biocca2002definingnetworkedminds}
\bibfield{author}{\bibinfo{person}{Frank Biocca} {and} \bibinfo{person}{Chad Harms}.} \bibinfo{year}{2002}\natexlab{}.
\newblock \showarticletitle{Defining and measuring social presence: Contribution to the networked minds theory and measure}.
\newblock \bibinfo{journal}{\emph{Proceedings of PRESENCE}}  \bibinfo{volume}{2002} (\bibinfo{year}{2002}), \bibinfo{pages}{7--36}.
\newblock


\bibitem[Bly et~al\mbox{.}(1993)]%
        {bly1993MediaSpacesBringing}
\bibfield{author}{\bibinfo{person}{Sara~A. Bly}, \bibinfo{person}{Steve~R. Harrison}, {and} \bibinfo{person}{Susan Irwin}.} \bibinfo{year}{1993}\natexlab{}.
\newblock \showarticletitle{Media spaces: bringing people together in a video, audio, and computing environment}.
\newblock \bibinfo{journal}{\emph{Commun. ACM}} \bibinfo{volume}{36}, \bibinfo{number}{1} (\bibinfo{date}{Jan.} \bibinfo{year}{1993}), \bibinfo{pages}{28--46}.
\newblock
\showISSN{0001-0782}
\href{https://doi.org/10.1145/151233.151235}{doi:\nolinkurl{10.1145/151233.151235}}


\bibitem[Braun and Clarke(2006)]%
        {braun2006UsingThematicAnalysis}
\bibfield{author}{\bibinfo{person}{Virginia Braun} {and} \bibinfo{person}{Victoria Clarke}.} \bibinfo{year}{2006}\natexlab{}.
\newblock \showarticletitle{Using thematic analysis in psychology}.
\newblock \bibinfo{journal}{\emph{Qualitative Research in Psychology}} \bibinfo{volume}{3}, \bibinfo{number}{2} (\bibinfo{date}{Jan.} \bibinfo{year}{2006}), \bibinfo{pages}{77--101}.
\newblock
\showISSN{1478-0887}
\href{https://doi.org/10.1191/1478088706qp063oa}{doi:\nolinkurl{10.1191/1478088706qp063oa}}


\bibitem[Braun and Clarke(2025)]%
        {braunvirginia2025ThematicAnalysisPractical}
\bibfield{author}{\bibinfo{person}{Virginia Braun} {and} \bibinfo{person}{Victoria Clarke}.} \bibinfo{year}{2025}\natexlab{}.
\newblock \bibinfo{title}{Thematic {Analysis}: {A} {Practical} {Guide}}.
\newblock
\urldef\tempurl%
\url{https://us.sagepub.com/en-us/nam/thematic-analysis/book248481}
\showURL{%
\tempurl}


\bibitem[Brubaker et~al\mbox{.}(2012)]%
        {brubaker2012FocusingSharedExperiences}
\bibfield{author}{\bibinfo{person}{Jed~R. Brubaker}, \bibinfo{person}{Gina Venolia}, {and} \bibinfo{person}{John~C. Tang}.} \bibinfo{year}{2012}\natexlab{}.
\newblock \showarticletitle{Focusing on shared experiences: moving beyond the camera in video communication}. In \bibinfo{booktitle}{\emph{Proceedings of the {Designing} {Interactive} {Systems} {Conference}}} \emph{(\bibinfo{series}{{DIS} '12})}. \bibinfo{publisher}{Association for Computing Machinery}, \bibinfo{address}{New York, NY, USA}, \bibinfo{pages}{96--105}.
\newblock
\showISBNx{978-1-4503-1210-3}
\href{https://doi.org/10.1145/2317956.2317973}{doi:\nolinkurl{10.1145/2317956.2317973}}


\bibitem[Buxton(2009)]%
        {buxton2009MediaspaceMeaningspaceMeetingspace}
\bibfield{author}{\bibinfo{person}{Bill Buxton}.} \bibinfo{year}{2009}\natexlab{}.
\newblock \showarticletitle{Mediaspace – {Meaningspace} – {Meetingspace}}.
\newblock In \bibinfo{booktitle}{\emph{Media {Space} 20 + {Years} of {Mediated} {Life}}}, \bibfield{editor}{\bibinfo{person}{Steve Harrison}} (Ed.). \bibinfo{publisher}{Springer}, \bibinfo{address}{London}, \bibinfo{pages}{217--231}.
\newblock
\showISBNx{978-1-84882-483-6}
\href{https://doi.org/10.1007/978-1-84882-483-6_13}{doi:\nolinkurl{10.1007/978-1-84882-483-6_13}}


\bibitem[Chatzopoulos et~al\mbox{.}(2017)]%
        {chatzopoulos2017MobileAugmentedReality}
\bibfield{author}{\bibinfo{person}{Dimitris Chatzopoulos}, \bibinfo{person}{Carlos Bermejo}, \bibinfo{person}{Zhanpeng Huang}, {and} \bibinfo{person}{Pan Hui}.} \bibinfo{year}{2017}\natexlab{}.
\newblock \showarticletitle{Mobile {Augmented} {Reality} {Survey}: {From} {Where} {We} {Are} to {Where} {We} {Go}}.
\newblock \bibinfo{journal}{\emph{IEEE Access}}  \bibinfo{volume}{5} (\bibinfo{year}{2017}), \bibinfo{pages}{6917--6950}.
\newblock
\showISSN{2169-3536}
\href{https://doi.org/10.1109/ACCESS.2017.2698164}{doi:\nolinkurl{10.1109/ACCESS.2017.2698164}}


\bibitem[Datcu et~al\mbox{.}(2016)]%
        {datcu2016HandheldAugmentedRealitya}
\bibfield{author}{\bibinfo{person}{Dragoş Datcu}, \bibinfo{person}{Stephan~G. Lukosch}, {and} \bibinfo{person}{Heide~K. Lukosch}.} \bibinfo{year}{2016}\natexlab{}.
\newblock \showarticletitle{Handheld {Augmented} {Reality} for {Distributed} {Collaborative} {Crime} {Scene} {Investigation}}. In \bibinfo{booktitle}{\emph{Proceedings of the 2016 {ACM} {International} {Conference} on {Supporting} {Group} {Work}}} \emph{(\bibinfo{series}{{GROUP} '16})}. \bibinfo{publisher}{Association for Computing Machinery}, \bibinfo{address}{New York, NY, USA}, \bibinfo{pages}{267--276}.
\newblock
\showISBNx{978-1-4503-4276-6}
\href{https://doi.org/10.1145/2957276.2957302}{doi:\nolinkurl{10.1145/2957276.2957302}}


\bibitem[Friard and Gamba(2016)]%
        {friard2016BORISFreeVersatile}
\bibfield{author}{\bibinfo{person}{Olivier Friard} {and} \bibinfo{person}{Marco Gamba}.} \bibinfo{year}{2016}\natexlab{}.
\newblock \showarticletitle{{BORIS}: a free, versatile open-source event-logging software for video/audio coding and live observations}.
\newblock \bibinfo{journal}{\emph{Methods in Ecology and Evolution}} \bibinfo{volume}{7}, \bibinfo{number}{11} (\bibinfo{year}{2016}), \bibinfo{pages}{1325--1330}.
\newblock
\showISSN{2041-210X}
\href{https://doi.org/10.1111/2041-210X.12584}{doi:\nolinkurl{10.1111/2041-210X.12584}}


\bibitem[Gauglitz et~al\mbox{.}(2012)]%
        {gauglitz2012IntegratingPhysicalEnvironment}
\bibfield{author}{\bibinfo{person}{Steffen Gauglitz}, \bibinfo{person}{Cha Lee}, \bibinfo{person}{Matthew Turk}, {and} \bibinfo{person}{Tobias Höllerer}.} \bibinfo{year}{2012}\natexlab{}.
\newblock \showarticletitle{Integrating the physical environment into mobile remote collaboration}. In \bibinfo{booktitle}{\emph{Proceedings of the 14th international conference on {Human}-computer interaction with mobile devices and services}} \emph{(\bibinfo{series}{{MobileHCI} '12})}. \bibinfo{publisher}{Association for Computing Machinery}, \bibinfo{address}{New York, NY, USA}, \bibinfo{pages}{241--250}.
\newblock
\showISBNx{978-1-4503-1105-2}
\href{https://doi.org/10.1145/2371574.2371610}{doi:\nolinkurl{10.1145/2371574.2371610}}


\bibitem[Gruenefeld et~al\mbox{.}(2018)]%
        {gruenefeld2018HaloWedgeVisualizing}
\bibfield{author}{\bibinfo{person}{Uwe Gruenefeld}, \bibinfo{person}{Abdallah~El Ali}, \bibinfo{person}{Susanne Boll}, {and} \bibinfo{person}{Wilko Heuten}.} \bibinfo{year}{2018}\natexlab{}.
\newblock \showarticletitle{Beyond {Halo} and {Wedge}: visualizing out-of-view objects on head-mounted virtual and augmented reality devices}. In \bibinfo{booktitle}{\emph{Proceedings of the 20th {International} {Conference} on {Human}-{Computer} {Interaction} with {Mobile} {Devices} and {Services}}} \emph{(\bibinfo{series}{{MobileHCI} '18})}. \bibinfo{publisher}{Association for Computing Machinery}, \bibinfo{address}{New York, NY, USA}, \bibinfo{pages}{1--11}.
\newblock
\showISBNx{978-1-4503-5898-9}
\href{https://doi.org/10.1145/3229434.3229438}{doi:\nolinkurl{10.1145/3229434.3229438}}


\bibitem[Gunkel et~al\mbox{.}(2021)]%
        {gunkel2021VRCommEndtoendWeb}
\bibfield{author}{\bibinfo{person}{Simon N.~B. Gunkel}, \bibinfo{person}{Rick Hindriks}, \bibinfo{person}{Karim M.~El Assal}, \bibinfo{person}{Hans~M. Stokking}, \bibinfo{person}{Sylvie Dijkstra-Soudarissanane}, \bibinfo{person}{Frank~ter Haar}, {and} \bibinfo{person}{Omar Niamut}.} \bibinfo{year}{2021}\natexlab{}.
\newblock \showarticletitle{{VRComm}: an end-to-end web system for real-time photorealistic social {VR} communication}. In \bibinfo{booktitle}{\emph{Proceedings of the 12th {ACM} {Multimedia} {Systems} {Conference}}} \emph{(\bibinfo{series}{{MMSys} '21})}. \bibinfo{publisher}{Association for Computing Machinery}, \bibinfo{address}{New York, NY, USA}, \bibinfo{pages}{65--79}.
\newblock
\showISBNx{978-1-4503-8434-6}
\href{https://doi.org/10.1145/3458305.3459595}{doi:\nolinkurl{10.1145/3458305.3459595}}


\bibitem[Gutwin and Greenberg(2002)]%
        {gutwin_descriptive_2002}
\bibfield{author}{\bibinfo{person}{Carl Gutwin} {and} \bibinfo{person}{Saul Greenberg}.} \bibinfo{year}{2002}\natexlab{}.
\newblock \showarticletitle{A {Descriptive} {Framework} of {Workspace} {Awareness} for {Real}-{Time} {Groupware}}.
\newblock \bibinfo{journal}{\emph{Computer Supported Cooperative Work (CSCW)}} \bibinfo{volume}{11}, \bibinfo{number}{3} (\bibinfo{date}{Sept.} \bibinfo{year}{2002}), \bibinfo{pages}{411--446}.
\newblock
\showISSN{1573-7551}
\href{https://doi.org/10.1023/A:1021271517844}{doi:\nolinkurl{10.1023/A:1021271517844}}


\bibitem[Heath and Luff(1991)]%
        {heath1991DisembodiedConductCommunication}
\bibfield{author}{\bibinfo{person}{Christian Heath} {and} \bibinfo{person}{Paul Luff}.} \bibinfo{year}{1991}\natexlab{}.
\newblock \showarticletitle{Disembodied conduct: communication through video in a multi-media office environment}. In \bibinfo{booktitle}{\emph{Proceedings of the {SIGCHI} {Conference} on {Human} {Factors} in {Computing} {Systems}}} \emph{(\bibinfo{series}{{CHI} '91})}. \bibinfo{publisher}{Association for Computing Machinery}, \bibinfo{address}{New York, NY, USA}, \bibinfo{pages}{99--103}.
\newblock
\showISBNx{978-0-89791-383-6}
\href{https://doi.org/10.1145/108844.108859}{doi:\nolinkurl{10.1145/108844.108859}}


\bibitem[Inkpen et~al\mbox{.}(2013)]%
        {inkpen2013Experiences2GoSharingKids}
\bibfield{author}{\bibinfo{person}{Kori Inkpen}, \bibinfo{person}{Brett Taylor}, \bibinfo{person}{Sasa Junuzovic}, \bibinfo{person}{John Tang}, {and} \bibinfo{person}{Gina Venolia}.} \bibinfo{year}{2013}\natexlab{}.
\newblock \showarticletitle{{Experiences2Go}: sharing kids' activities outside the home with remote family members}. In \bibinfo{booktitle}{\emph{Proceedings of the 2013 conference on {Computer} supported cooperative work}} \emph{(\bibinfo{series}{{CSCW} '13})}. \bibinfo{publisher}{Association for Computing Machinery}, \bibinfo{address}{New York, NY, USA}, \bibinfo{pages}{1329--1340}.
\newblock
\showISBNx{978-1-4503-1331-5}
\href{https://doi.org/10.1145/2441776.2441926}{doi:\nolinkurl{10.1145/2441776.2441926}}


\bibitem[Irlitti et~al\mbox{.}(2023)]%
        {irlitti2023VolumetricMixedReality}
\bibfield{author}{\bibinfo{person}{Andrew Irlitti}, \bibinfo{person}{Mesut Latifoglu}, \bibinfo{person}{Qiushi Zhou}, \bibinfo{person}{Martin~N Reinoso}, \bibinfo{person}{Thuong Hoang}, \bibinfo{person}{Eduardo Velloso}, {and} \bibinfo{person}{Frank Vetere}.} \bibinfo{year}{2023}\natexlab{}.
\newblock \showarticletitle{Volumetric {Mixed} {Reality} {Telepresence} for {Real}-time {Cross} {Modality} {Collaboration}}. In \bibinfo{booktitle}{\emph{Proceedings of the 2023 {CHI} {Conference} on {Human} {Factors} in {Computing} {Systems}}} \emph{(\bibinfo{series}{{CHI} '23})}. \bibinfo{publisher}{Association for Computing Machinery}, \bibinfo{address}{New York, NY, USA}, \bibinfo{pages}{1--14}.
\newblock
\showISBNx{978-1-4503-9421-5}
\href{https://doi.org/10.1145/3544548.3581277}{doi:\nolinkurl{10.1145/3544548.3581277}}


\bibitem[Ishii and Miyake(1991)]%
        {ishii1991OpenSharedWorkspace}
\bibfield{author}{\bibinfo{person}{Hiroshi Ishii} {and} \bibinfo{person}{Naomi Miyake}.} \bibinfo{year}{1991}\natexlab{}.
\newblock \showarticletitle{Toward an open shared workspace: computer and video fusion approach of {TeamWorkStation}}.
\newblock \bibinfo{journal}{\emph{Commun. ACM}} \bibinfo{volume}{34}, \bibinfo{number}{12} (\bibinfo{date}{Dec.} \bibinfo{year}{1991}), \bibinfo{pages}{37--50}.
\newblock
\showISSN{0001-0782}
\href{https://doi.org/10.1145/125319.125321}{doi:\nolinkurl{10.1145/125319.125321}}


\bibitem[Jones et~al\mbox{.}(2022)]%
        {jones2022RescueCASTRExploringPhotos}
\bibfield{author}{\bibinfo{person}{Brennan Jones}, \bibinfo{person}{Anthony Tang}, {and} \bibinfo{person}{Carman Neustaedter}.} \bibinfo{year}{2022}\natexlab{}.
\newblock \showarticletitle{{RescueCASTR}: {Exploring} {Photos} and {Live} {Streaming} to {Support} {Contextual} {Awareness} in the {Wilderness} {Search} and {Rescue} {Command} {Post}}.
\newblock \bibinfo{journal}{\emph{Proc. ACM Hum.-Comput. Interact.}} \bibinfo{volume}{6}, \bibinfo{number}{CSCW1} (\bibinfo{date}{April} \bibinfo{year}{2022}), \bibinfo{pages}{113:1--113:32}.
\newblock
\href{https://doi.org/10.1145/3512960}{doi:\nolinkurl{10.1145/3512960}}


\bibitem[Jones et~al\mbox{.}(2015)]%
        {jones2015MechanicsCameraWork}
\bibfield{author}{\bibinfo{person}{Brennan Jones}, \bibinfo{person}{Anna Witcraft}, \bibinfo{person}{Scott Bateman}, \bibinfo{person}{Carman Neustaedter}, {and} \bibinfo{person}{Anthony Tang}.} \bibinfo{year}{2015}\natexlab{}.
\newblock \showarticletitle{Mechanics of {Camera} {Work} in {Mobile} {Video} {Collaboration}}. In \bibinfo{booktitle}{\emph{Proceedings of the 33rd {Annual} {ACM} {Conference} on {Human} {Factors} in {Computing} {Systems}}} \emph{(\bibinfo{series}{{CHI} '15})}. \bibinfo{publisher}{Association for Computing Machinery}, \bibinfo{address}{New York, NY, USA}, \bibinfo{pages}{957--966}.
\newblock
\showISBNx{978-1-4503-3145-6}
\href{https://doi.org/10.1145/2702123.2702345}{doi:\nolinkurl{10.1145/2702123.2702345}}


\bibitem[Kim et~al\mbox{.}(2018)]%
        {kim2018EffectCollaborationStyles}
\bibfield{author}{\bibinfo{person}{Seungwon Kim}, \bibinfo{person}{Mark Billinghurst}, {and} \bibinfo{person}{Gun Lee}.} \bibinfo{year}{2018}\natexlab{}.
\newblock \showarticletitle{The {Effect} of {Collaboration} {Styles} and {View} {Independence} on {Video}-{Mediated} {Remote} {Collaboration}}.
\newblock \bibinfo{journal}{\emph{Computer Supported Cooperative Work (CSCW)}} \bibinfo{volume}{27}, \bibinfo{number}{3} (\bibinfo{date}{Dec.} \bibinfo{year}{2018}), \bibinfo{pages}{569--607}.
\newblock
\showISSN{1573-7551}
\href{https://doi.org/10.1007/s10606-018-9324-2}{doi:\nolinkurl{10.1007/s10606-018-9324-2}}


\bibitem[Klemmer et~al\mbox{.}(2006)]%
        {klemmer2006HowBodiesMattera}
\bibfield{author}{\bibinfo{person}{Scott~R. Klemmer}, \bibinfo{person}{Björn Hartmann}, {and} \bibinfo{person}{Leila Takayama}.} \bibinfo{year}{2006}\natexlab{}.
\newblock \showarticletitle{How bodies matter: five themes for interaction design}. In \bibinfo{booktitle}{\emph{Proceedings of the 6th conference on {Designing} {Interactive} systems}} \emph{(\bibinfo{series}{{DIS} '06})}. \bibinfo{publisher}{Association for Computing Machinery}, \bibinfo{address}{New York, NY, USA}, \bibinfo{pages}{140--149}.
\newblock
\showISBNx{978-1-59593-367-6}
\href{https://doi.org/10.1145/1142405.1142429}{doi:\nolinkurl{10.1145/1142405.1142429}}


\bibitem[Kuckartz and Rädiker(2019)]%
        {kuckartz2019AnalyzingQualitativeData}
\bibfield{author}{\bibinfo{person}{Udo Kuckartz} {and} \bibinfo{person}{Stefan Rädiker}.} \bibinfo{year}{2019}\natexlab{}.
\newblock \bibinfo{booktitle}{\emph{Analyzing {Qualitative} {Data} with {MAXQDA}: {Text}, {Audio}, and {Video}}}.
\newblock \bibinfo{publisher}{Springer International Publishing}, \bibinfo{address}{Cham}.
\newblock
\showISBNx{978-3-030-15670-1 978-3-030-15671-8}
\href{https://doi.org/10.1007/978-3-030-15671-8}{doi:\nolinkurl{10.1007/978-3-030-15671-8}}


\bibitem[Kuzuoka(1992)]%
        {kuzuoka1992SpatialWorkspaceCollaboration}
\bibfield{author}{\bibinfo{person}{Hideaki Kuzuoka}.} \bibinfo{year}{1992}\natexlab{}.
\newblock \showarticletitle{Spatial workspace collaboration: a {SharedView} video support system for remote collaboration capability}. In \bibinfo{booktitle}{\emph{Proceedings of the {SIGCHI} {Conference} on {Human} {Factors} in {Computing} {Systems}}} \emph{(\bibinfo{series}{{CHI} '92})}. \bibinfo{publisher}{Association for Computing Machinery}, \bibinfo{address}{New York, NY, USA}, \bibinfo{pages}{533--540}.
\newblock
\showISBNx{978-0-89791-513-7}
\href{https://doi.org/10.1145/142750.142980}{doi:\nolinkurl{10.1145/142750.142980}}


\bibitem[Li et~al\mbox{.}(2022)]%
        {li2022ARCritiqueSupportingRemote}
\bibfield{author}{\bibinfo{person}{Yuan Li}, \bibinfo{person}{Sang~Won Lee}, \bibinfo{person}{Doug~A. Bowman}, \bibinfo{person}{David Hicks}, \bibinfo{person}{Wallace~santos Lages}, {and} \bibinfo{person}{Akshay Sharma}.} \bibinfo{year}{2022}\natexlab{}.
\newblock \showarticletitle{{ARCritique}: {Supporting} {Remote} {Design} {Critique} of {Physical} {Artifacts} through {Collaborative} {Augmented} {Reality}}. In \bibinfo{booktitle}{\emph{Proceedings of the 2022 {ACM} {Symposium} on {Spatial} {User} {Interaction}}} \emph{(\bibinfo{series}{{SUI} '22})}. \bibinfo{publisher}{Association for Computing Machinery}, \bibinfo{address}{New York, NY, USA}, \bibinfo{pages}{1--12}.
\newblock
\showISBNx{978-1-4503-9948-7}
\href{https://doi.org/10.1145/3565970.3567700}{doi:\nolinkurl{10.1145/3565970.3567700}}


\bibitem[Licoppe and Morel(2009)]%
        {licoppe2009CollaborativeWorkProducing}
\bibfield{author}{\bibinfo{person}{Christian Licoppe} {and} \bibinfo{person}{Julien Morel}.} \bibinfo{year}{2009}\natexlab{}.
\newblock \showarticletitle{The collaborative work of producing meaningful shots in mobile video telephony}. In \bibinfo{booktitle}{\emph{Proceedings of the 11th {International} {Conference} on {Human}-{Computer} {Interaction} with {Mobile} {Devices} and {Services}}} \emph{(\bibinfo{series}{{MobileHCI} '09})}. \bibinfo{publisher}{Association for Computing Machinery}, \bibinfo{address}{New York, NY, USA}, \bibinfo{pages}{1--10}.
\newblock
\showISBNx{978-1-60558-281-8}
\href{https://doi.org/10.1145/1613858.1613903}{doi:\nolinkurl{10.1145/1613858.1613903}}


\bibitem[Luff and Heath(1998)]%
        {luff1998MobilityCollaboration}
\bibfield{author}{\bibinfo{person}{Paul Luff} {and} \bibinfo{person}{Christian Heath}.} \bibinfo{year}{1998}\natexlab{}.
\newblock \showarticletitle{Mobility in collaboration}. In \bibinfo{booktitle}{\emph{Proceedings of the 1998 {ACM} {Conference} on {Computer} {Supported} {Cooperative} {Work}}} \emph{(\bibinfo{series}{{CSCW} '98})}. \bibinfo{publisher}{Association for Computing Machinery}, \bibinfo{address}{New York, NY, USA}, \bibinfo{pages}{305--314}.
\newblock
\showISBNx{978-1-58113-009-6}
\href{https://doi.org/10.1145/289444.289505}{doi:\nolinkurl{10.1145/289444.289505}}


\bibitem[Luff et~al\mbox{.}(2003)]%
        {luff2003FracturedEcologiesCreating}
\bibfield{author}{\bibinfo{person}{Paul Luff}, \bibinfo{person}{Christian Heath}, \bibinfo{person}{Hideaki Kuzuoka}, \bibinfo{person}{Jon Hindmarsh}, \bibinfo{person}{Keiichi Yamazaki}, {and} \bibinfo{person}{Shinya Oyama}.} \bibinfo{year}{2003}\natexlab{}.
\newblock \showarticletitle{Fractured {Ecologies}: {Creating} {Environments} for {Collaboration}}.
\newblock \bibinfo{journal}{\emph{Human–Computer Interaction}} \bibinfo{volume}{18}, \bibinfo{number}{1-2} (\bibinfo{date}{June} \bibinfo{year}{2003}), \bibinfo{pages}{51--84}.
\newblock
\showISSN{0737-0024}
\href{https://doi.org/10.1207/S15327051HCI1812_3}{doi:\nolinkurl{10.1207/S15327051HCI1812_3}}


\bibitem[Mackay and McGrenere(2025)]%
        {mackay2025ComparativeStructuredObservation}
\bibfield{author}{\bibinfo{person}{Wendy~E. Mackay} {and} \bibinfo{person}{Joanna McGrenere}.} \bibinfo{year}{2025}\natexlab{}.
\newblock \showarticletitle{Comparative Structured Observation}.
\newblock \bibinfo{journal}{\emph{ACM Trans. Comput.-Hum. Interact.}} \bibinfo{volume}{32}, \bibinfo{number}{2}, Article \bibinfo{articleno}{14} (\bibinfo{date}{April} \bibinfo{year}{2025}), \bibinfo{numpages}{27}~pages.
\newblock
\showISSN{1073-0516}
\href{https://doi.org/10.1145/3711838}{doi:\nolinkurl{10.1145/3711838}}


\bibitem[Mantei et~al\mbox{.}(1991)]%
        {mantei1991ExperiencesUseMedia}
\bibfield{author}{\bibinfo{person}{Marilyn~M. Mantei}, \bibinfo{person}{Ronald~M. Baecker}, \bibinfo{person}{Abigail~J. Sellen}, \bibinfo{person}{William A.~S. Buxton}, \bibinfo{person}{Thomas Milligan}, {and} \bibinfo{person}{Barry Wellman}.} \bibinfo{year}{1991}\natexlab{}.
\newblock \showarticletitle{Experiences in the use of a media space}. In \bibinfo{booktitle}{\emph{Proceedings of the {SIGCHI} {Conference} on {Human} {Factors} in {Computing} {Systems}}} \emph{(\bibinfo{series}{{CHI} '91})}. \bibinfo{publisher}{Association for Computing Machinery}, \bibinfo{address}{New York, NY, USA}, \bibinfo{pages}{203--208}.
\newblock
\showISBNx{978-0-89791-383-6}
\href{https://doi.org/10.1145/108844.108888}{doi:\nolinkurl{10.1145/108844.108888}}


\bibitem[Müller et~al\mbox{.}(2017)]%
        {muller2017RemoteCollaborationMixed}
\bibfield{author}{\bibinfo{person}{Jens Müller}, \bibinfo{person}{Roman Rädle}, {and} \bibinfo{person}{Harald Reiterer}.} \bibinfo{year}{2017}\natexlab{}.
\newblock \showarticletitle{Remote {Collaboration} {With} {Mixed} {Reality} {Displays}: {How} {Shared} {Virtual} {Landmarks} {Facilitate} {Spatial} {Referencing}}. In \bibinfo{booktitle}{\emph{Proceedings of the 2017 {CHI} {Conference} on {Human} {Factors} in {Computing} {Systems}}} \emph{(\bibinfo{series}{{CHI} '17})}. \bibinfo{publisher}{Association for Computing Machinery}, \bibinfo{address}{New York, NY, USA}, \bibinfo{pages}{6481--6486}.
\newblock
\showISBNx{978-1-4503-4655-9}
\href{https://doi.org/10.1145/3025453.3025717}{doi:\nolinkurl{10.1145/3025453.3025717}}


\bibitem[Müller et~al\mbox{.}(2019)]%
        {muller2019QualitativeComparisonAugmenteda}
\bibfield{author}{\bibinfo{person}{Jens Müller}, \bibinfo{person}{Johannes Zagermann}, \bibinfo{person}{Jonathan Wieland}, \bibinfo{person}{Ulrike Pfeil}, {and} \bibinfo{person}{Harald Reiterer}.} \bibinfo{year}{2019}\natexlab{}.
\newblock \showarticletitle{A {Qualitative} {Comparison} {Between} {Augmented} and {Virtual} {Reality} {Collaboration} with {Handheld} {Devices}}. In \bibinfo{booktitle}{\emph{Proceedings of {Mensch} und {Computer} 2019}} \emph{(\bibinfo{series}{{MuC} '19})}. \bibinfo{publisher}{Association for Computing Machinery}, \bibinfo{address}{New York, NY, USA}, \bibinfo{pages}{399--410}.
\newblock
\showISBNx{978-1-4503-7198-8}
\href{https://doi.org/10.1145/3340764.3340773}{doi:\nolinkurl{10.1145/3340764.3340773}}


\bibitem[O'Hara et~al\mbox{.}(2006)]%
        {ohara2006EverydayPracticesMobile}
\bibfield{author}{\bibinfo{person}{Kenton O'Hara}, \bibinfo{person}{Alison Black}, {and} \bibinfo{person}{Matthew Lipson}.} \bibinfo{year}{2006}\natexlab{}.
\newblock \showarticletitle{Everyday practices with mobile video telephony}. In \bibinfo{booktitle}{\emph{Proceedings of the {SIGCHI} {Conference} on {Human} {Factors} in {Computing} {Systems}}} \emph{(\bibinfo{series}{{CHI} '06})}. \bibinfo{publisher}{Association for Computing Machinery}, \bibinfo{address}{New York, NY, USA}, \bibinfo{pages}{871--880}.
\newblock
\showISBNx{978-1-59593-372-0}
\href{https://doi.org/10.1145/1124772.1124900}{doi:\nolinkurl{10.1145/1124772.1124900}}


\bibitem[Orts-Escolano et~al\mbox{.}(2016)]%
        {orts-escolano2016HoloportationVirtual3D}
\bibfield{author}{\bibinfo{person}{Sergio Orts-Escolano}, \bibinfo{person}{Christoph Rhemann}, \bibinfo{person}{Sean Fanello}, \bibinfo{person}{Wayne Chang}, \bibinfo{person}{Adarsh Kowdle}, \bibinfo{person}{Yury Degtyarev}, \bibinfo{person}{David Kim}, \bibinfo{person}{Philip~L. Davidson}, \bibinfo{person}{Sameh Khamis}, \bibinfo{person}{Mingsong Dou}, \bibinfo{person}{Vladimir Tankovich}, \bibinfo{person}{Charles Loop}, \bibinfo{person}{Qin Cai}, \bibinfo{person}{Philip~A. Chou}, \bibinfo{person}{Sarah Mennicken}, \bibinfo{person}{Julien Valentin}, \bibinfo{person}{Vivek Pradeep}, \bibinfo{person}{Shenlong Wang}, \bibinfo{person}{Sing~Bing Kang}, \bibinfo{person}{Pushmeet Kohli}, \bibinfo{person}{Yuliya Lutchyn}, \bibinfo{person}{Cem Keskin}, {and} \bibinfo{person}{Shahram Izadi}.} \bibinfo{year}{2016}\natexlab{}.
\newblock \showarticletitle{Holoportation: {Virtual} {3D} {Teleportation} in {Real}-time}. In \bibinfo{booktitle}{\emph{Proceedings of the 29th {Annual} {Symposium} on {User} {Interface} {Software} and {Technology}}} \emph{(\bibinfo{series}{{UIST} '16})}. \bibinfo{publisher}{Association for Computing Machinery}, \bibinfo{address}{New York, NY, USA}, \bibinfo{pages}{741--754}.
\newblock
\showISBNx{978-1-4503-4189-9}
\href{https://doi.org/10.1145/2984511.2984517}{doi:\nolinkurl{10.1145/2984511.2984517}}


\bibitem[Piumsomboon et~al\mbox{.}(2019)]%
        {piumsomboon2019ShoulderGiantMultiScale}
\bibfield{author}{\bibinfo{person}{Thammathip Piumsomboon}, \bibinfo{person}{Gun~A. Lee}, \bibinfo{person}{Andrew Irlitti}, \bibinfo{person}{Barrett Ens}, \bibinfo{person}{Bruce~H. Thomas}, {and} \bibinfo{person}{Mark Billinghurst}.} \bibinfo{year}{2019}\natexlab{}.
\newblock \showarticletitle{On the {Shoulder} of the {Giant}: {A} {Multi}-{Scale} {Mixed} {Reality} {Collaboration} with 360 {Video} {Sharing} and {Tangible} {Interaction}}. In \bibinfo{booktitle}{\emph{Proceedings of the 2019 {CHI} {Conference} on {Human} {Factors} in {Computing} {Systems}}} \emph{(\bibinfo{series}{{CHI} '19})}. \bibinfo{publisher}{Association for Computing Machinery}, \bibinfo{address}{New York, NY, USA}, \bibinfo{pages}{1--17}.
\newblock
\showISBNx{978-1-4503-5970-2}
\href{https://doi.org/10.1145/3290605.3300458}{doi:\nolinkurl{10.1145/3290605.3300458}}


\bibitem[Poelman et~al\mbox{.}(2012)]%
        {poelman2012IfBeingThere}
\bibfield{author}{\bibinfo{person}{Ronald Poelman}, \bibinfo{person}{Oytun Akman}, \bibinfo{person}{Stephan Lukosch}, {and} \bibinfo{person}{Pieter Jonker}.} \bibinfo{year}{2012}\natexlab{}.
\newblock \showarticletitle{As if being there: mediated reality for crime scene investigation}. In \bibinfo{booktitle}{\emph{Proceedings of the {ACM} 2012 conference on {Computer} {Supported} {Cooperative} {Work}}} \emph{(\bibinfo{series}{{CSCW} '12})}. \bibinfo{publisher}{Association for Computing Machinery}, \bibinfo{address}{New York, NY, USA}, \bibinfo{pages}{1267--1276}.
\newblock
\showISBNx{978-1-4503-1086-4}
\href{https://doi.org/10.1145/2145204.2145394}{doi:\nolinkurl{10.1145/2145204.2145394}}


\bibitem[Procyk et~al\mbox{.}(2014)]%
        {procyk2014ExploringVideoStreaming}
\bibfield{author}{\bibinfo{person}{Jason Procyk}, \bibinfo{person}{Carman Neustaedter}, \bibinfo{person}{Carolyn Pang}, \bibinfo{person}{Anthony Tang}, {and} \bibinfo{person}{Tejinder~K. Judge}.} \bibinfo{year}{2014}\natexlab{}.
\newblock \showarticletitle{Exploring video streaming in public settings: shared geocaching over distance using mobile video chat}. In \bibinfo{booktitle}{\emph{Proceedings of the {SIGCHI} {Conference} on {Human} {Factors} in {Computing} {Systems}}} \emph{(\bibinfo{series}{{CHI} '14})}. \bibinfo{publisher}{Association for Computing Machinery}, \bibinfo{address}{New York, NY, USA}, \bibinfo{pages}{2163--2172}.
\newblock
\showISBNx{978-1-4503-2473-1}
\href{https://doi.org/10.1145/2556288.2557198}{doi:\nolinkurl{10.1145/2556288.2557198}}


\bibitem[Sellen(1995)]%
        {sellen1995RemoteConversationsEffects}
\bibfield{author}{\bibinfo{person}{Abigail~J. Sellen}.} \bibinfo{year}{1995}\natexlab{}.
\newblock \showarticletitle{Remote {Conversations}: {The} {Effects} of {Mediating} {Talk} {With} {Technology}}.
\newblock \bibinfo{journal}{\emph{Human–Computer Interaction}} \bibinfo{volume}{10}, \bibinfo{number}{4} (\bibinfo{date}{Dec.} \bibinfo{year}{1995}), \bibinfo{pages}{401--444}.
\newblock
\showISSN{0737-0024}
\href{https://doi.org/10.1207/s15327051hci1004_2}{doi:\nolinkurl{10.1207/s15327051hci1004_2}}


\bibitem[Shakeri et~al\mbox{.}(2017)]%
        {shakeri2017EscapingTogetherDesign}
\bibfield{author}{\bibinfo{person}{Hanieh Shakeri}, \bibinfo{person}{Samarth Singhal}, \bibinfo{person}{Rui Pan}, \bibinfo{person}{Carman Neustaedter}, {and} \bibinfo{person}{Anthony Tang}.} \bibinfo{year}{2017}\natexlab{}.
\newblock \showarticletitle{Escaping {Together}: {The} {Design} and {Evaluation} of a {Distributed} {Real}-{Life} {Escape} {Room}}. In \bibinfo{booktitle}{\emph{Proceedings of the {Annual} {Symposium} on {Computer}-{Human} {Interaction} in {Play}}} \emph{(\bibinfo{series}{{CHI} {PLAY} '17})}. \bibinfo{publisher}{Association for Computing Machinery}, \bibinfo{address}{New York, NY, USA}, \bibinfo{pages}{115--128}.
\newblock
\showISBNx{978-1-4503-4898-0}
\href{https://doi.org/10.1145/3116595.3116601}{doi:\nolinkurl{10.1145/3116595.3116601}}


\bibitem[Stachniss et~al\mbox{.}(2016)]%
        {stachniss2016SimultaneousLocalizationMapping}
\bibfield{author}{\bibinfo{person}{Cyrill Stachniss}, \bibinfo{person}{John~J. Leonard}, {and} \bibinfo{person}{Sebastian Thrun}.} \bibinfo{year}{2016}\natexlab{}.
\newblock \showarticletitle{Simultaneous {Localization} and {Mapping}}.
\newblock In \bibinfo{booktitle}{\emph{Springer {Handbook} of {Robotics}}}, \bibfield{editor}{\bibinfo{person}{Bruno Siciliano} {and} \bibinfo{person}{Oussama Khatib}} (Eds.). \bibinfo{publisher}{Springer International Publishing}, \bibinfo{address}{Cham}, \bibinfo{pages}{1153--1176}.
\newblock
\showISBNx{978-3-319-32552-1}
\href{https://doi.org/10.1007/978-3-319-32552-1_46}{doi:\nolinkurl{10.1007/978-3-319-32552-1_46}}


\bibitem[Thoravi~Kumaravel et~al\mbox{.}(2019)]%
        {thoravikumaravel2019LokiFacilitatingRemote}
\bibfield{author}{\bibinfo{person}{Balasaravanan Thoravi~Kumaravel}, \bibinfo{person}{Fraser Anderson}, \bibinfo{person}{George Fitzmaurice}, \bibinfo{person}{Bjoern Hartmann}, {and} \bibinfo{person}{Tovi Grossman}.} \bibinfo{year}{2019}\natexlab{}.
\newblock \showarticletitle{Loki: {Facilitating} {Remote} {Instruction} of {Physical} {Tasks} {Using} {Bi}-{Directional} {Mixed}-{Reality} {Telepresence}}. In \bibinfo{booktitle}{\emph{Proceedings of the 32nd {Annual} {ACM} {Symposium} on {User} {Interface} {Software} and {Technology}}} \emph{(\bibinfo{series}{{UIST} '19})}. \bibinfo{publisher}{Association for Computing Machinery}, \bibinfo{address}{New York, NY, USA}, \bibinfo{pages}{161--174}.
\newblock
\showISBNx{978-1-4503-6816-2}
\href{https://doi.org/10.1145/3332165.3347872}{doi:\nolinkurl{10.1145/3332165.3347872}}


\bibitem[Tran et~al\mbox{.}(2024)]%
        {tran_classifying_2024}
\bibfield{author}{\bibinfo{person}{Tanh~Quang Tran}, \bibinfo{person}{Tobias Langlotz}, \bibinfo{person}{Jacob Young}, \bibinfo{person}{Thomas~W. Schubert}, {and} \bibinfo{person}{Holger Regenbrecht}.} \bibinfo{year}{2024}\natexlab{}.
\newblock \showarticletitle{Classifying {Presence} {Scores}: {Insights} and {Analysis} from {Two} {Decades} of the {Igroup} {Presence} {Questionnaire} ({IPQ})}.
\newblock \bibinfo{journal}{\emph{ACM Trans. Comput.-Hum. Interact.}} \bibinfo{volume}{31}, \bibinfo{number}{5} (\bibinfo{date}{Nov.} \bibinfo{year}{2024}), \bibinfo{pages}{61:1--61:26}.
\newblock
\showISSN{1073-0516}
\href{https://doi.org/10.1145/3689046}{doi:\nolinkurl{10.1145/3689046}}


\bibitem[Unver et~al\mbox{.}(2018)]%
        {unver2018HandsFreeRemoteCollaboration}
\bibfield{author}{\bibinfo{person}{Baris Unver}, \bibinfo{person}{Sarah D'Angelo}, \bibinfo{person}{Matthew Miller}, \bibinfo{person}{John~C. Tang}, \bibinfo{person}{Gina Venolia}, {and} \bibinfo{person}{Kori Inkpen}.} \bibinfo{year}{2018}\natexlab{}.
\newblock \showarticletitle{Hands-{Free} {Remote} {Collaboration} {Over} {Video}: {Exploring} {Viewer} and {Streamer} {Reactions}}. In \bibinfo{booktitle}{\emph{Proceedings of the 2018 {ACM} {International} {Conference} on {Interactive} {Surfaces} and {Spaces}}} \emph{(\bibinfo{series}{{ISS} '18})}. \bibinfo{publisher}{Association for Computing Machinery}, \bibinfo{address}{New York, NY, USA}, \bibinfo{pages}{85--95}.
\newblock
\showISBNx{978-1-4503-5694-7}
\href{https://doi.org/10.1145/3279778.3279803}{doi:\nolinkurl{10.1145/3279778.3279803}}


\bibitem[Vanukuru et~al\mbox{.}(2023)]%
        {vanukuru2023DualStreamSpatiallySharing}
\bibfield{author}{\bibinfo{person}{Rishi Vanukuru}, \bibinfo{person}{Suibi Che-Chuan Weng}, \bibinfo{person}{Krithik Ranjan}, \bibinfo{person}{Torin Hopkins}, \bibinfo{person}{Amy Banic}, \bibinfo{person}{Mark~D. Gross}, {and} \bibinfo{person}{Ellen Yi-Luen Do}.} \bibinfo{year}{2023}\natexlab{}.
\newblock \showarticletitle{{DualStream}: {Spatially} {Sharing} {Selves} and {Surroundings} using {Mobile} {Devices} and {Augmented} {Reality}}. In \bibinfo{booktitle}{\emph{2023 IEEE International Symposium on Mixed and Augmented Reality (ISMAR)}}. \bibinfo{pages}{138--147}.
\newblock
\href{https://doi.org/10.1109/ISMAR59233.2023.00028}{doi:\nolinkurl{10.1109/ISMAR59233.2023.00028}}


\bibitem[Villanueva et~al\mbox{.}(2022)]%
        {villanueva2022ColabARToolkitRemotea}
\bibfield{author}{\bibinfo{person}{Ana Villanueva}, \bibinfo{person}{Zhengzhe Zhu}, \bibinfo{person}{Ziyi Liu}, \bibinfo{person}{Feiyang Wang}, \bibinfo{person}{Subramanian Chidambaram}, {and} \bibinfo{person}{Karthik Ramani}.} \bibinfo{year}{2022}\natexlab{}.
\newblock \showarticletitle{{ColabAR}: {A} {Toolkit} for {Remote} {Collaboration} in {Tangible} {Augmented} {Reality} {Laboratories}}.
\newblock \bibinfo{journal}{\emph{Proc. ACM Hum.-Comput. Interact.}} \bibinfo{volume}{6}, \bibinfo{number}{CSCW1} (\bibinfo{date}{April} \bibinfo{year}{2022}), \bibinfo{pages}{81:1--81:22}.
\newblock
\href{https://doi.org/10.1145/3512928}{doi:\nolinkurl{10.1145/3512928}}


\bibitem[Young et~al\mbox{.}(2020)]%
        {young2020MobileportationNomadicTelepresence}
\bibfield{author}{\bibinfo{person}{Jacob Young}, \bibinfo{person}{Tobias Langlotz}, \bibinfo{person}{Steven Mills}, {and} \bibinfo{person}{Holger Regenbrecht}.} \bibinfo{year}{2020}\natexlab{}.
\newblock \showarticletitle{Mobileportation: {Nomadic} {Telepresence} for {Mobile} {Devices}}.
\newblock \bibinfo{journal}{\emph{Proc. ACM Interact. Mob. Wearable Ubiquitous Technol.}} \bibinfo{volume}{4}, \bibinfo{number}{2} (\bibinfo{date}{June} \bibinfo{year}{2020}), \bibinfo{pages}{65:1--65:16}.
\newblock
\href{https://doi.org/10.1145/3397331}{doi:\nolinkurl{10.1145/3397331}}


\end{thebibliography}



\newpage
\appendix

\section{Quantitative Results}

\begin{table}[ht]
\centering
\caption{Statistical overview of spatial presence and recall scores across the four conditions of Onsite Spatial, Onsite Video, Remote Spatial, and Remote Video}
\label{tab:quant-table}
\begin{tabular}{|l|l|lll|l|lll|}
\hline
                            & \multirow{6}{*}{} & \multicolumn{3}{c|}{\textit{\textbf{Presence (1-7)}}}                                        & \multirow{6}{*}{} & \multicolumn{3}{c|}{\textit{\textbf{Recall}}}                                          \\ \cline{1-1} \cline{3-5} \cline{7-9} 
\textit{\textbf{Condition}} &                   & \multicolumn{1}{l|}{\textbf{Mean}} & \multicolumn{1}{l|}{\textbf{SD}} & \textbf{Range} &                   & \multicolumn{1}{l|}{\textbf{Mean}} & \multicolumn{1}{l|}{\textbf{SD}} & \textbf{Range} \\ \cline{1-1} \cline{3-5} \cline{7-9} 
\textbf{Onsite Spatial}     &                   & \multicolumn{1}{l|}{4.87}          & \multicolumn{1}{l|}{0.68}        & (3.9, 6.0)     &                   & \multicolumn{1}{l|}{27.71}         & \multicolumn{1}{l|}{7.01}        & (16, 38)       \\ \cline{1-1} \cline{3-5} \cline{7-9} 
\textbf{Onsite Video}       &                   & \multicolumn{1}{l|}{3.84}          & \multicolumn{1}{l|}{1.02}        & (2.3, 5.5)     &                   & \multicolumn{1}{l|}{27.29}         & \multicolumn{1}{l|}{6.09}        & (16, 37)       \\ \cline{1-1} \cline{3-5} \cline{7-9} 
\textbf{Remote Spatial}     &                   & \multicolumn{1}{l|}{4.65}          & \multicolumn{1}{l|}{0.76}        & (3.6, 5.7)     &                   & \multicolumn{1}{l|}{26.71}         & \multicolumn{1}{l|}{7.39}        & (13, 39)       \\ \cline{1-1} \cline{3-5} \cline{7-9} 
\textbf{Remote Video}       &                   & \multicolumn{1}{l|}{3.33}          & \multicolumn{1}{l|}{1.06}        & (1.7, 5.6)     &                   & \multicolumn{1}{l|}{21.79}         & \multicolumn{1}{l|}{7.33}        & (4, 32)        \\ \hline
\end{tabular}
\end{table}

\subsection{Items from the Presence Questionnaire}

We used a subset of questions from the igroup Presence Questionnaire \cite{tran_classifying_2024} and the Networked Minds Presence Inventory \cite{biocca2002definingnetworkedminds}. All questions were rated on a Likert scale from 1 to 7.

\begin{enumerate}
    \item How much did it seem as if the objects and people you saw/heard had come to the place you were? (Not at all --- Very much)
    \item How often when a person seemed to be headed toward you did you want to move to get out of its way? (Never --- Always)
    \item To what extent did you experience a sense of being there inside the environment you saw? (Not at all --- Very much)
    \item Did the experience seem more like looking at the events/people on a movie screen or more like looking at the events/people through a window? (Like a movie screen --- Like a window)
    \item How often did you have the sensation that people you saw/heard could also see/hear you? (Never --- Always)
    \item To what extent did you feel you could interact with the person or people you saw? (None --- Very much)
    \item How much did it seem as if you and the people you saw were together in the same place? (Not at all --- Very much)
    \item How often did it feel as if someone you saw/heard in the environment was talking directly to you? (Never --- Always)
    \item How often did you want to or did you make eye-contact with someone you saw? (Never --- Always)
    \item During the experience how well were you able to observe the facial expressions of the person you saw? (Not well --- Very well)
\end{enumerate}

\section{Additional Figures}

\begin{figure}[ht]
  \centering
  \includegraphics[width=\linewidth]{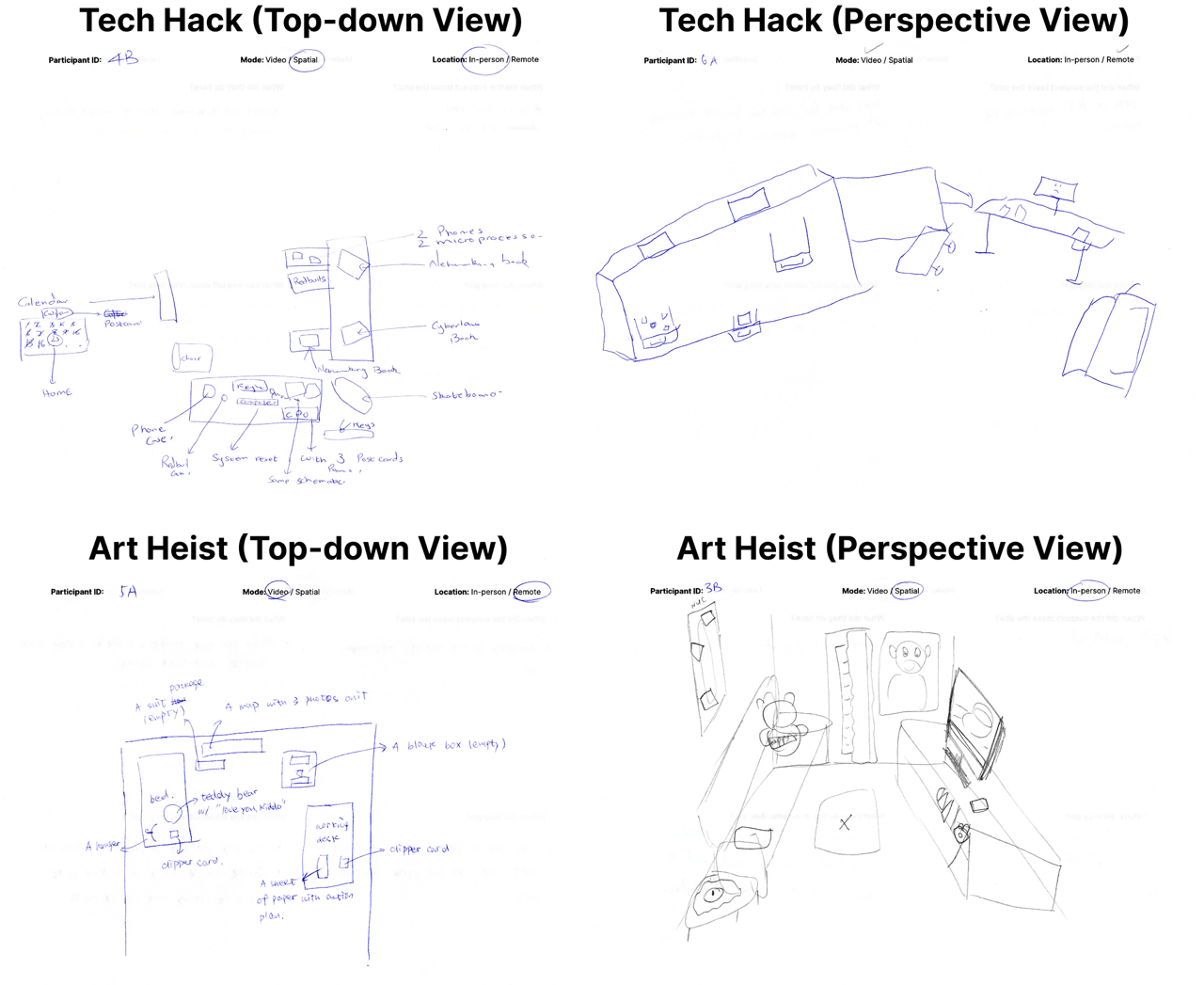}
  \caption{Hand-drawn maps from the post-task worksheet by the participants. The figure shows two types of maps, top-down and perspective, for each of the Art Heist and Tech Hack scenes.}
  \label{fig:drawn-maps}
  \Description{This figure contains four scans of the participants' hand drawn maps, two for each scene, with one in top-down projection and one in perspective view each. The top down view sketches contain labels for objects, while the perspective sketches do not.}
\end{figure}

\begin{figure}[h]
  \includegraphics[width=\textwidth]{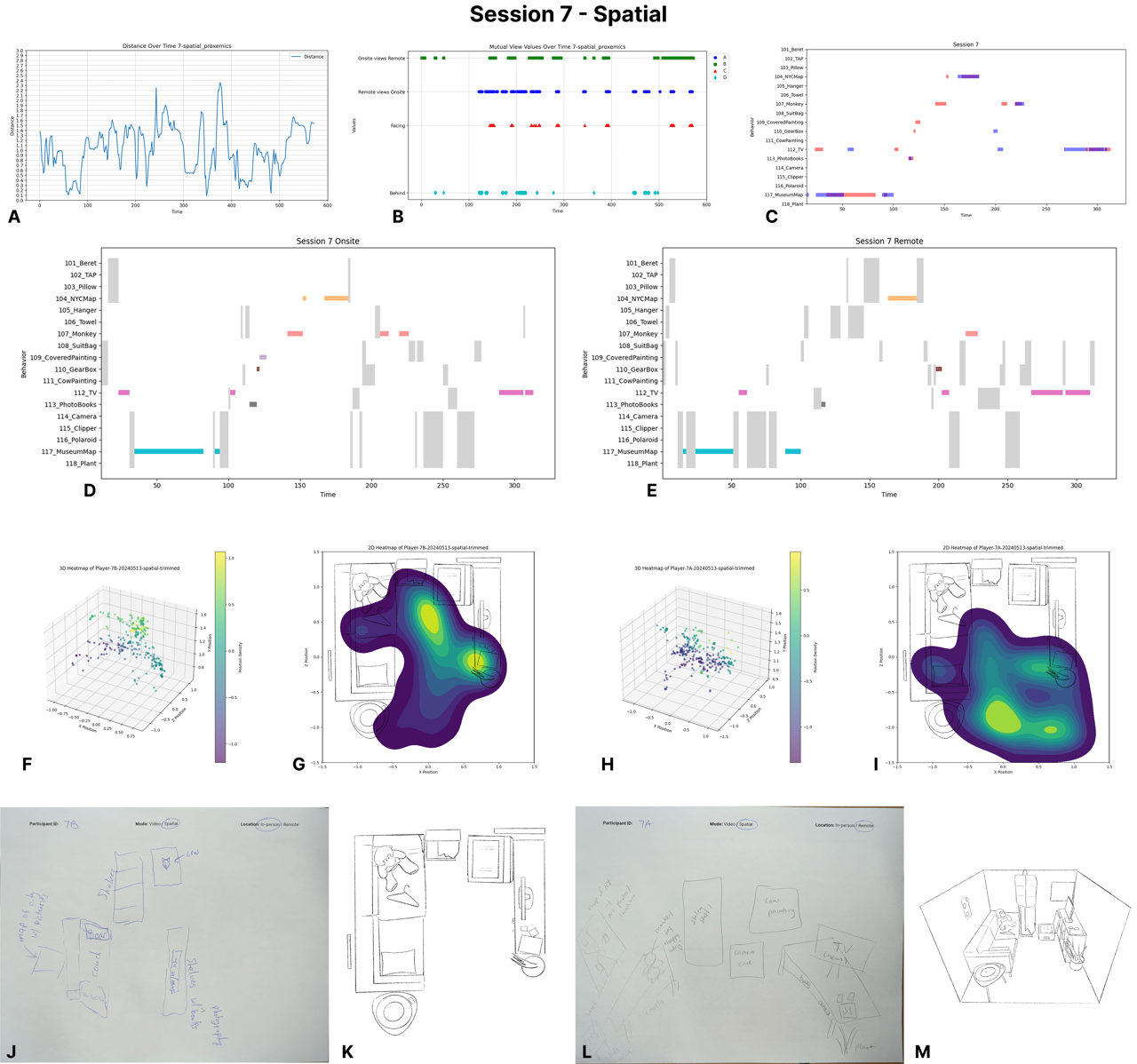}
  \caption{Session Map constructed for Session 7 in spatial mode in the Art Heist scene. (A) Distance between the two participants in space, derived using in-app motion logs. (B) Spatial proxemics over time, categorized by whether (i) OP views RP's representation, (ii) RP views OP's representation, (iii) both view each other, or (iv) one is behind the other. (C-E) Object focus during the call of (D) OP, (E) RP, and (C) overlaid. (F-I) 2D and 3D heatmaps of movements of OP (F-G) and RP (H-I) derived using motion logs. (J-M) Maps drawn by the participants post task completion, compared with actual maps: (J-K) for OP, (L-M) for RP.}
  \label{fig:session-spatial}
  \Description{This is a composite figure containing multiple plots and maps relating the Session 7 AR.}
\end{figure}

\textbf{\begin{figure}[h]
  \includegraphics[width=\textwidth]{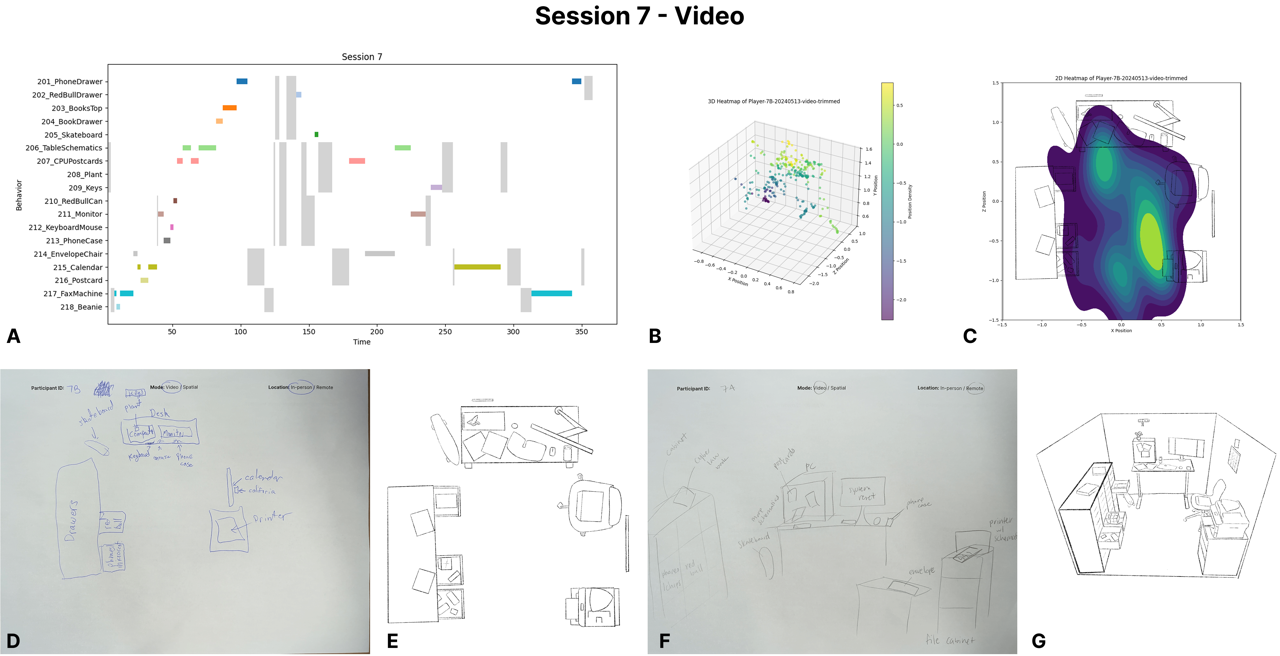}
  \caption{Session Map constructed for Session 7 in video mode in the Tech Hack scene. (A) Object focus of the video call. (B-C) 2D and 3D heatmaps of OP derived using motion logs. (D-G) Maps drawn by the participants post task completion, compared with actual maps: (D-E) for OP, (F-G) for RP.}
  \label{fig:session-video}
  \Description{This is a composite figure containing multiple plots and maps relating the Session 7 video.}
\end{figure}}

\end{document}